\documentclass[12pt,onecolumn]{IEEEtran}
\usepackage{amssymb,amsmath,mathrsfs,verbatim,bm,epsfig,color,cite,accents}
\usepackage{hyperref}

\input{epic.sty}
\input{eepic.sty}

\input{psfig}               
\input{epsf}                
\renewcommand{\baselinestretch}{1.5}
\renewcommand{\IEEEQED}{\IEEEQEDopen}
\newtheorem{thm}{Theorem}[section]
\newtheorem{cor}[thm]{Corollary}
\newtheorem{lem}[thm]{Lemma}

\newcommand{\bthm} {\begin{thm} }
\newcommand{\ethm} {\end{thm}}
\newcommand{\blem} {\begin{lem} }
\newcommand{\elem} {\end{lem}}
\newcommand{\bcor} {\begin{cor} }
\newcommand{\ecor} {\end{cor}}

\newcommand{\beq}  {\begin{equation}}
\newcommand{\eeq}  {\end{equation}}
\newcommand{\beqa}{\begin{eqnarray}}
\newcommand{\eeqa} {\end{eqnarray}}
\newcommand{\bdis} {\begin{displaymath}}
\newcommand{\edis} {\end{displaymath}}
\newcommand{\bitem}{\begin{itemize}}
\newcommand{\eitem}{\end{itemize}}
\newcommand{\ba}  {\begin{array}}
\newcommand{\ea}  {\end{array}}

\newcommand{\err}  {{\rm err}}
\newcommand{\ber}  {{\rm BER}}

\newcommand{\Th}   {{\rm Th}}
\newcommand{\av}   {{\rm av}}
\newcommand{\low}   {{\rm L}}
\newcommand{\up}   {{\rm U}}
\newcommand{\eve}   {{\rm e}}
\newcommand{\bob}   {{\rm b}}

\newcommand{\sigb}   {{\sigma^2_{\rm b}}}
\newcommand{\sige}   {{\sigma^2_{\rm e}}}

\newcommand{\non}      {\nonumber}

\newcommand{\opt}      {{\rm opt}}

\newcommand{\calP}   {{\cal P}}

\newcommand{\calC}   {{\cal C}}

\newcommand{\calR}   {{\cal R}}

\newcommand{\bfq} {{\bf q}}

\begin{document}

\begin{center}
\vspace*{15mm}
\Large BER-Based Physical Layer Security with Finite Codelength: Combining Strong  Converse and  Error Amplification
\vspace{20mm} \normalsize \\
Il-Min Kim, Byoung-Hoon Kim, and Joon Kui Ahn \\
\vspace{5mm}
Department of Electrical and Computer Engineering \\
Queen's University \\
Kingston, ON, Canada   K7M 2A8 \\
E-mail: ilmin.kim@queensu.ca \\
\vspace{20mm}
January 4, 2015
\end{center}

\newpage

\begin{abstract}
A bit error rate (BER)-based physical layer security approach is proposed for finite blocklength. For secure communication in the sense of high BER, the information-theoretic strong converse is combined with cryptographic error amplification achieved by substitution permutation networks (SPNs) based on confusion and diffusion. For discrete memoryless channels (DMCs), an analytical framework is provided showing the tradeoffs among finite blocklength, maximum/minimum possible transmission rates, and BER requirements for the legitimate receiver and the eavesdropper. Also, the security gap is analytically studied for Gaussian channels and the concept is extended to other DMCs including binary symmetric channels (BSCs) and binary erasure channels (BECs). For fading channels, the transmit power is optimized to minimize the outage probability of the legitimate receiver subject to a BER threshold for the eavesdropper.
\end{abstract}


\section{Introduction}
Security is a critical issue in communications \cite{shannon} and it is particularly challenging with a growing number of different wireless communication applications and various wireless devices. Due to the broadcast nature of wireless medium, the wireless security is inherently more vulnerable than the wired security: the eavesdropper may overhear and interpret the messages in wireless communications more easily than in wireline communications. Traditionally, the issue of security has been addressed at a higher layer by cryptography, which requires secret keys. A problem of this approach is that it is often challenging to distribute and manage the secret keys, especially for many emerging wireless networks. Furthermore, once the devices are physically compromised by an adversary, the communication is no longer secure.

As a fundamentally different approach, the physical layer security, particularly information theoretic security, has received a lot of attention. The information-theoretic security is based on the pioneering work of \cite{wyner}, where the channel from the transmitter (Alice) to the eavesdropper (Eve) was assumed to be a degraded version of the channel from Alice to the legitimate receiver (Bob), namely the degraded wiretap channel. For this channel, Wyner derived the capacity-equivocation region. Later, this work was extended to the non-degraded case, where the eavesdropper's channel is not necessarily a degraded version of the legitimate user's channel \cite{csiszar1}, and also applied to Gaussian channels \cite{leung}. Recently, the information theoretic security and/or the physical layer security have regained much interest for secure wireless communications.

In most of the works in the area of physical layer security, the security metric is defined based on mutual information between Alice and Eve. If the mutual information is strictly zero, it is perfectly secure, called perfect security \cite{shannon}. With perfect security, Eve cannot obtain any additional information about Alice's message from Eve's received signal. However, in order to ensure such perfect security, the entropy of the secret key must not be smaller than the entropy of the source message. In real communications, therefore, it is not practical to try to achieve the perfect security. Addressing this issue, two non-perfect security notions have been extensively studied: weak secrecy \cite{wyner} and strong secrecy. The weak secrecy requires that the mutual information rate, i.e., mutual information divided by the blocklength (or codelength), approaches zero when the blocklength goes to infinity. On the other hand, the strong secrecy requires that the mutual information itself approaches zero when the blocklength goes to infinity. Many researchers have designed codes providing the weak secrecy or strong secrecy \cite{thangaraj}--\cite{bloch}. Vast majority of the works have been devoted to the weak secrecy, mostly based on low-density parity-check (LDPC) codes \cite{thangaraj}, \cite{rathi} or polar codes \cite{mahdavifar} (also, see the references in \cite{bloch}). Because designing codes to achieve the strong secrecy is generally much more difficult, the works for the strong secrecy were generally limited to simplistic scenarios such as noiseless Bob's channel \cite{mahdavifar}, \cite{Subramanian} or binary symmetric channel \cite{cheraghchi}. However, it has been argued that the weak secrecy might be a too weak security condition \cite{Maurer1, Maurer2}, and in fact, one can easily construct examples of codes achieving weak security that are never secure in practice \cite{bloch}. A problem of those codes in \cite{thangaraj}--\cite{bloch} is that they are not directly applicable to continuous-input channels, such as additive white Gaussian noise (AWGN) or fading channels. Another (perhaps more serious) problem is that, for finite blocklength, it is not clear how to evaluate or quantify the strength of security actually achieved by the codes designed based on strong or weak secrecy secrecy. Unless the blocklength is very long, the codes might not be secure enough to be used in practical systems, especially for the case of weak secrecy.

Other than the information theoretic security notions based on mutual information, there are few other security measures considered in the literature. For example, the signal-to-interference-plus-noise ratio (SINR) has been used as a secrecy measure in the area of physical layer security based on signal processing techniques \cite{W_C_Liao}, \cite{Yang}. However, it is unclear how to exactly set an SINR threshold and to evaluate what strength (or kind) of security can be actually achieved by an SNIR threshold. Another security measure considered in the literature is the block error probability, i.e., decoding error probability of codeword. When the transmission rate is above the channel capacity, by the strong converse \cite{Arimoto}, the block error probability approaches one as the blocklength tends to infinity. Given a security condition in terms of block error probability, for a Gaussian wiretap channel, the authors of \cite{Rajesh} studied the asymptotic transmission rate and the rate for finite blocklength using a rate approximation expression \cite[Theorem 54]{polyanskiy} for AWGN channels. In \cite{Belfiore}, a coset lattice code was designed to ensure high block error probability at Eve and low block error probability at Bob. However, a limitation of the approach based on the block error probability only  is that high block error probability at Eve does not necessarily mean secure communication. This is because a block error event simply means that there is at least one bit error within a block (or codeword). As an example, if there is always only one bit error in a block, the block error probability is one. However, all the remaining bits except the particular single bit can be decoded by Eve, which is certainly not secure.

Arguably, a {\it practically} effective and useful security measure in the physical layer security might be the bit error rate (BER). If it is possible to ensure that Eve's BER is (very close to) 0.5, she essentially cannot recover any information bits transmitted by Alice. In \cite{d_klinc}, for AWGN channels, punctured LDPC codes were designed to ensure high BER at Eve. The analysis was limited to asymptotic case of LDPC codes and Eve's BER is evaluated only by simulations, from which it is not easy to obtain any theoretical insights. In \cite{M_Baldi}, to induce high BER at Eve for AWGN channels, Bose-Chaudhuri-Hocuenghem (BCH) codes and LDPC codes are combined with scrambling/descrambling. The BER analysis of BCH codes was based on an approximate BER equation of \cite{D_Torrieri} under the assumption of bounded-distance decoding with hard decision, and the study on LDPC codes was purely based on simulations.

In this paper, we also adopt the BER as the security measure for Eve. Using Gallager's random coding exponent and the strong converse over general discrete memoryless channels (DMCs), we first ensure that Bob's block error probability tends to zero and Eve's block error probability tends to one. To amplify the errors such that Eve's BER is close to 0.5, we then utilize substitution permutation networks (SPNs). In particular, the error amplification by SPN is not only mathematically analyzed based on the ideal modeling, but also numerically evaluated based on actual simulation of a real SPN. Given BER requirements for Bob and Eve, for finite blocklength, we analyze the maximum and minimum possible transmission rates.  Also, the security gap is defined and analyzed for AWGN channels and then the concept is  extended to other DMCs. Focusing on Gaussian-input fading channels, we analytically optimize the transmit power to minimize Bob's reliability outage probability, subject to a security condition given in terms of a BER lower-bound threshold for Eve. The summary of the contributions is as follows:
\bitem
\item For secure communication in the sense of high BER, the information-theoretic strong converse is combined with cryptographic error amplification achieved by SPNs based on confusion and diffusion.
\item For DMCs, an analytical framework is provided showing the trade-offs among finite blocklength, maximum/minimum possible transmission rates, and BER requirements for Bob and Eve.
\item For Gaussian channels, with finite blocklength, the security gap is analytically studied and the concept is extended to other DMCs including binary symmetric channels (BSCs) and binary erasure channels (BECs).
\item For fading channels, with finite blocklength, the transmit power is analytically optimized to minimize Bob's outage probability subject to a BER threshold for Eve.
\eitem

A practical benefit of the BER-based physical layer security is particularly evident when both Bob's and Eve's channels are good and the channel quality difference is small: $C_\bob > C_\eve \gg 1$ with $C_\bob- C_\eve \ll 1$, where $C_\bob$ is Bob's capacity and $C_\eve$ is Eve's capacity. If the weak secrecy or strong secrecy constraint is imposed, the transmission rate is bounded by the secrecy capacity given by $C_\bob-C_\eve \ll 1$ for the channels such as symmetric degraded wiretap channels \cite{leung_2} or Gaussian channels \cite{leung}. On the other hand, if the high BER condition is imposed as a security constraint and our approach is taken, the transmission rate can go up to $C_\bob \gg 1$. Another benefit of the proposed approach is that, for finite blocklength, we can ensure a high target BER requirement for Eve, whereas for weak/strong secrecy, it is not entirely clear how to ensure a particular security requirement with finite blocklength.

The rest of this paper is organized as follows. In Section II, Gallager's random coding exponent and the strong converse are reviewed to derive Bob's block error probability upper-bound and Eve's block error probability lower-bound. Also, it is demonstrated that the errors can be effectively amplified by SPNs. In Section III, we first combine the strong converse and the SPNs. Then the maximum/minimum rates and security gaps are analyzed given finite blocklength and the BER requirements for Bob and Eve. Also, for fading channels, the transmission power is optimized to minimize the reliability outage probability subject to a security condition. In Section IV, some numerical results are presented and the paper is concluded in Section V.

{\it Notation:} We use A := B to denote that A, by definition, is equal to B,
and we use A =: B to denote that B, by definition, is equal
to A. Also, ${\cal CN}(0,\sigma^2)$ denotes a circularly symmetric complex Gaussian distribution with variance $\sigma^2$ (or variance $\sigma^2/2$ per dimension).

%


\section{Gallger Function, Strong Converse, and Error Amplification}


Assume that message $M$ represented by $K$ bits is transmitted by Alice. Using a code composed of $2^K$ codewords, the message is encoded into a codeword $X^n$ of $n$ symbols. The transmission rate $R$ is given by \beq R=\frac{K \ln 2}{n} ~~~~ {\rm (nats/channel ~ use)}.\eeq Bob's received codeword is denoted by $Y_\bob^n$ and Eve's received codeword is denoted by $Y_\eve^n$. Assuming both channels are DMCs, they are described by the conditional probability distributions $f_{Y_\bob|X}(y_\bob|x)$ and $f_{Y_\eve|X}(y_\eve|x)$, respectively, for Bob and Eve. Let $\hat {M_\bob}$ and $\hat {M_\eve}$ denote the decoded messages at Bob and Eve, respectively. Let $C_\bob$ and $C_\eve$ denote the channel capacities for Bob and Eve, respectively.

\subsection{Bob's Block Error Probability based on Gallager Function}
Let $\calC$ denote a code whose symbols $X$ are randomly generated by input distribution $q_X(x)$, which is simply denoted by $q(x)$ whenever there is no ambiguity. Let $P^\bob_\err(R|\calC)=\Pr(M \neq \hat M_\bob|\calC)$ denote the decoding error probability of code $\calC$ at Bob. Let $P^\bob_\err(R)$ denote the average probability  over the ensemble of all codes at Bob. The ensemble average block error probability $P^\bob_\err(R)$ at Bob can be upper-bounded as follows \cite[Theorem 5.6.2]{gallager}:
\beq P^\bob_\err(R)=\mathbb{E}[P^\bob_\err(R|\calC)] \leq P_\err^{\bob, \up}(R,\rho, q(x)) \label{eq:upper_bound_Bob}  \eeq where the upper-bound $P_\err^{\bob, \up}(R,\rho, q(x))$ is given by
\beqa P_\err^{\bob,\up}(R,\rho, q(x)) &=& \exp \left( -n  \left\{  E_0^\bob(\rho, q(x)) - \rho R \right\} \right), ~~~ 0 \leq \rho \leq 1. \label{eq:P_err_bob_init} \eeqa In the above equation, Gallager function $E_0^\bob(\rho, q(x))$ is given by \beqa E_0^\bob(\rho, q(x)) = -\ln \sum_{y_\bob} \left[ \sum_x q(x) f_{Y_\bob|X}(y_\bob|x)^{\frac{1}{1+\rho}}   \right]^{1+\rho}, ~~~ 0 \leq \rho \leq 1 \label{eq:E_0_bob} \eeqa where $\sum_x$ is replaced by $\int_x$ if $X$ is continuous, and $\sum_{Y_\bob}$ is replaced by $\int_{Y_\bob}$ if $Y_\bob$ is continuous.
Since the upper-bound $P_\err^{\bob, \up}(R,\rho, q(x))$ is valid for any $0 \leq \rho \leq 1$ and for any distribution $q(x)$, the bound can be tightened by optimizing $\rho$ and $q(x)$ as follows:
\beqa \min_{0 \leq  \rho \leq 1}  \min_{q(x)}P_\err^{\bob, \up}(R,\rho, q(x))  \eeqa or \beqa  \min_{0 \leq  \rho \leq 1}   \left\{ \max_{q(x)}  E_0^\bob(\rho, q(x)) - \rho R \right\}. \eeqa In this paper, we will use $\breve q(x)$ and $\breve \rho$ to denote the optimal distribution and optimal $\rho$, respectively, which are defined as follows: \beqa \breve q(x) &=& \arg \min_{q(x)} P_\err^{\bob, \up}(R,\rho, q(x))=\arg \max_{q(x)} E_0^\bob(\rho, q(x)) \\ \breve \rho &=& \arg \min_{0 \leq \rho \leq 1} P_\err^{\bob, \up}(R,\rho, \breve q(x))=\arg \max_{0 \leq \rho \leq 1}\left\{  E_0^\bob(\rho, \breve q(x)) - \rho R \right\}. \eeqa

When $R< I_\bob(q(x))$, the exponent in (\ref{eq:P_err_bob_init}) is positive with maximization over $\rho$ \cite[Section 5.6, p. 143]{gallager}:
\beqa  \max_{0 \leq  \rho \leq 1} \left\{  E_0^\bob(\rho , q(x))- \rho R \right\} & >&  0, ~~~ R <  I_\bob(q(x)) . \eeqa
When $R< C_\bob$, the exponent in (\ref{eq:P_err_bob_init}) is positive with maximization over $q(x)$ and $\rho$ \cite[Section 5.6, p. 143]{gallager}:
\beqa  \max_{0 \leq  \rho \leq 1} \left\{ \max_{q(x)} E_0^\bob(\rho , q(x))- \rho R \right\} & >&  0, ~~~ R <  C_\bob . \eeqa
When $R< C_\bob$, therefore, there exists at least one code of which block error probability upper-bound tends exponentially to zero as $n \rightarrow \infty$. With the optimal $\breve q(x)$ yielding the tightest upper-bound, the asymptotic slope of $E_0^\bob(\rho,q(x))$ when $\rho$ approaches zero from the right is the capacity of Bob's channel
\cite[Section 5.6]{gallager}: \beqa C_\bob  &=&  \lim_{\rho \downarrow 0} \frac{1}{\rho} \max_{q(x)} E_0^\bob(\rho, q(x)) \label{eq:asympt_rho_1} \\ &=& \max_{q(x)} \left. \frac{\partial}{\partial \rho} E_0^\bob(\rho, q(x)) \right|_{\rho=0}. \label{eq:asympt_rho_2}  \eeqa

\subsection{Eve's Block Error Probability based on Arimoto's Strong Converse}
Let $P^\eve_\err(R|\calC)=\Pr(M \neq \hat M_\eve|\calC)$ denote the block error probability of code $\calC$ at Eve. We first define $P_\err^{\eve,\low}(R,\rho', q'(x))$ as follows \beqa P_\err^{\eve,\low}(R,\rho',q'(x))  &=&  1-\exp \left( -n \left\{   E^\eve_0(\rho', q'(x)) - \rho' R \right\} \right), ~~~ -1< \rho' \leq 0 \label{eq:P_err_eve_init} \eeqa where $E_0^\eve(\rho', q'(x))$ is given by (\ref{eq:E_0_bob}) with $q(x)$, $f_{Y_\bob|x}(y_\bob|x)$, and $\rho$ replaced by $q'(x)$, $f_{Y_\eve|x}(y_\eve|x)$, and $\rho'$, respectively. When $R>C_\eve$ and a priori probabilities are equal, Eve's block error probability $P^\eve_\err(R|\calC)$ of {\it any} code $\calC$ is be lower bounded by \cite{Arimoto}, \cite[Eq. (3.9.21)]{viterbi}:
\beq P^\eve_\err(R|\calC) \geq   P_\err^{\eve,\low}(R,\rho', \breve q'(x)), ~~~ \forall \calC \label{eq:lower_bound_Eve}  \eeq
where $\breve q'(x)$ is given by \beq \breve q'(x) = \arg \min_{q'(x)}  P_\err^{\eve,\low}(R,\rho',q'(x)) =\arg \min_{q'(x)}  E^\eve_0(\rho', q'(x)). \label{eq:def_breve_q_prime} \eeq
Note that, unlike the case of upper-bound, the single-letter expression (\ref{eq:lower_bound_Eve}) of the lower-bound is obtained with the particular input distribution $\breve q'(x)$.\footnote{Compared to the upper-bound tightened by $\breve q(x)$, the lower-bound determined by $\breve q'(x)$ might be considered to be weaker or less tight because $\breve q'(x)$ is obtained by minimizing $E_0^\eve(\rho',q(x))$ rather than maximizing it. In return, the obtained lower-bound is valid for all possible codes (rather than some codes in the ensemble as in the upper-bound case).}
Since lower-bound $P_\err^{\eve, \low}(R, \rho', \breve q'(x))$ is still valid
for any $-1 < \rho' \leq 0$, the tightest bound can be obtained by optimizing $\rho'$ as follows: \beqa \breve \rho' &=& \arg \max_{-1 < \rho' \leq 0} P_\err^{\eve, \low}(R,\rho', \breve q'(x))=\arg \max_{-1 < \rho' \leq 0}\left\{  E_0^\eve(\rho', \breve q'(x)) - \rho' R \right\}. \eeqa

When $R> C_\eve$, the exponent in (\ref{eq:P_err_eve_init}) is positive with maximization over $\rho'$ and minimization over $q'(x)$: \cite[Theorem 2]{Arimoto}\cite[Theorem 3.9.1]{viterbi}:
\beqa  \max_{-1 < \rho' \leq 0} \left\{ \min_{q'(x)}  E^\eve_0(\rho', q'(x)) - \rho' R \right\}  >  0, ~~~ R >  C_\eve . \eeqa
When $R > C_\eve$ and a priori probabilities are equal, therefore, the error probability upper-bound of any code tends exponentially to one as $n \rightarrow \infty$.
With the particular distribution $\breve q'(x)$ yielding the valid lower-bound for any code, the asymptotic slope of $E_0^\eve(\rho',q'(x))$ when $\rho'$ approaches zero from the left is the capacity of Eve's channel \cite{Arimoto}: \beqa C_\eve &=&  \lim_{\rho' \uparrow 0} \frac{1}{\rho'} \min_{q'(x)} E_0^\eve(\rho', q'(x)) \label{eq:asympt_rho_prime_1}  \\ &=& \max_{q'(x)} \left. \frac{\partial}{\partial \rho'} E_0^\eve(\rho',q'(x)) \right|_{\rho'=0} . \label{eq:asympt_rho_prime_2} \eeqa

\subsection{Confusion and Diffusion: Error Amplification by SPN in Cryptography}
In this subsection, the issue of error amplification is discussed.
In cryptography, error amplification has been extensively and systematically studied for various applications including hash functions and block ciphers such as Data Encryption Standard (DES) and Advanced Encryption Standard (AES) \cite{Stinson_book}. A most common approach is to use substitution-boxes (S-boxes), which are designed based on several criterions such as the completeness, avalanche property, etc. In particular, the avalanche property plays a very important role. This property was first introduced by Feistel \cite{Feistel}; but, the fundamental concept was actually based on Shannon's confusion \cite{shannon}. In \cite{webster_1}, strict avalanche criterion (SAC) was defined as follows: SAC is satisfied if, whenever a single input bit is complemented, each of all output bits changes with a 50\% probability.
Also, high degree SAC can be defined \cite{Meier}--\cite{cusick_book}: SAC of degree $l$ is satisfied if, whenever $l$ input bits are complemented at the same time, each of the output bits changes with a 50\% probability.

In general, it is very difficult to design  large-size S-boxes satisfying SAC. In today's practical cryptographic systems, therefore, small-size S-boxes are often used; for example, $8 \times 8$ S-boxes are used for AES.  In order to handle a larger number of input bits at the same time, substitution-permutation networks (SPNs) are often used. An SPN is composed of multiple parallel-connected S-boxes taking multiple input bits. The output bits from those S-boxes are permutated by a permutation box (P-box). Typically, an SPN is designed by implementing several rounds of alternating S-boxes and P-boxes.\footnote{For example, AES has 10 rounds for 128 bit secret keys, 12 rounds for 192 bit secret keys, and 14 rounds for 256 bit secret keys.} In fact, the design of alternating S-boxes and P-boxes is based on Shannon's two fundamental security concepts: confusion and diffusion \cite{shannon}. In SPNs for cryptographic applications, secret keys are typically used. In this paper, however, we do not use any secret keys for SPNs because we will use SPNs only to amplify the errors (rather than encrypting data as in cryptography). In the following, the error amplification effect of SPNs is evaluated first by analysis assuming ideal S-boxes and then by simulation using real S-boxes.

In \cite{Hey_1}, assuming ideal S-boxes satisfying SAC, the output error probability of the SPN  was analyzed. Let $K$ denote the number of input and output bits of the SPN. Let $W_r$ denote the random variable representing the number of bit errors after round $r$. Let $B$ denote the number of input and output bits of each S-box. Assuming $K$ is an integer multiples of $B$, we use $J = \frac{K}{B}$ to denote the number of S-boxes connected in parallel for each round.  Let $L_r$ denote the random variable representing the number of S-boxes in round $r$ affected by the bit errors.
The distribution of $W_r$ is given by \cite{Hey_1}\footnote{Although this expression is given in closed-form, it becomes difficult to use as $K$ increases, because the computational complexity grows with $K$ very quickly.}
\beqa q_{W_r}(w_r) = \sum_{l_r=1}^J f_{W_r|L_r}(w_r|l_r) \sum_{w_{r-1}=1}^K f_{L_r|W_{r-1}}( l_r| w_{r-1} )q_{W_{r-1}}(w_{r-1}), {\rm ~ for~} w_r=1,\cdots, K \label{eq:distribution_W_r} \eeqa where \beqa f_{L_r|W_{r-1}}(l_r|w_{r-1}) &=& \frac{A_1(l_r,w_{r-1})}{A_2(w_{r-1})}, {\rm ~ for ~} l_r=1,\cdots, J \\ A_1(l,w) &=& \sum_{i=J-l}^J (-1)^{i-(J-l)}{i \choose J-l}{J \choose i} { (J-i)B \choose w}^+ \\ A_2(w) &=& {K \choose w} \\ f_{W_r|L_r}(w_r|l_r) &=& \frac{1}{(2^B-1)^{l_r}}\sum_{i=0}^{l_r} (-1)^i { l_r \choose i} { (l_r-i)B \choose w_r}^+ .   \eeqa
In the above equation, ${a \choose b}^+={a \choose b}$ if $a \geq b$;  ${a \choose b}^+=0$ if $a < b$. Using $q_{W_r}(w_r)$, the BER at the output of the SPN after $r$ rounds can be determined as follows\beq   P_{\ber}^{\rm SPN}(r,K)  = \frac{1}{K}\sum_{w_r=1}^K w_r q_{W_r}(w_r), ~~~~ r =1,2,\cdots. \label{eq:BER_SPN_ana} \eeq
In order to actually determine the BER using (\ref{eq:BER_SPN_ana}), the initial distribution $q_{W_0}(w_0)$ must be explicitly given. As an example, for the scenario where there is only a single input bit error, the initial distribution is given by
\beq q_{W_0} (w_0)= \left\{ \begin{array}{ll} 1,& {\rm if~} w_0 =1 \\ 0,& {\rm otherwise}. \end{array} \right. \label{eq:initial_distribution} \eeq

\begin{figure}[h]
\begin{center}
\includegraphics[width=0.7\columnwidth]{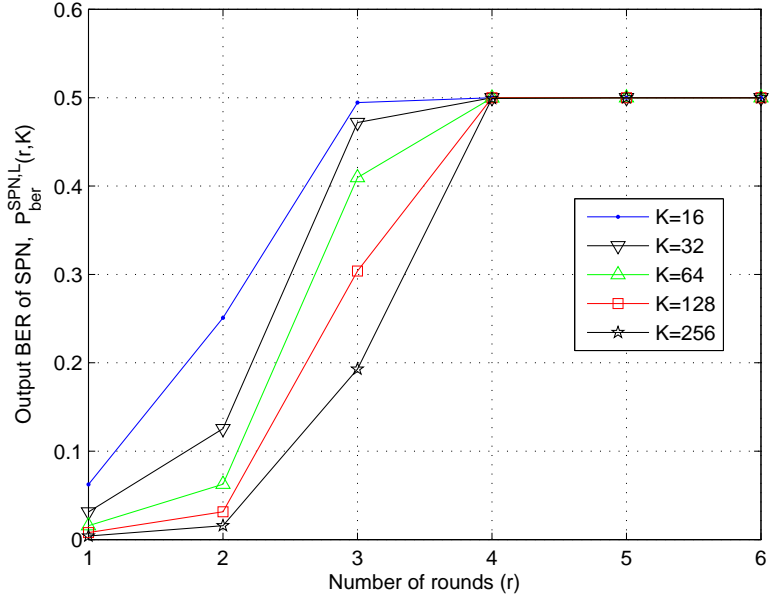}
\caption{BER at the output of the SPN composed of $8 \times 8$ {\it theoretical} S-boxes satisfying SAC, when only one input bit is in error out of total $K$ input bits. The number $J$ of S-boxes for each round is given by $\frac{K}{8}$. The BER is analytically obtained by (\ref{eq:distribution_W_r})--(\ref{eq:initial_distribution}).}
\label{fig:SPN_BER_ana}
\end{center}
\end{figure}

In Fig. \ref{fig:SPN_BER_ana}, the output BER analytically obtained by (\ref{eq:distribution_W_r})--(\ref{eq:initial_distribution}) is plotted for different sizes of SPNs with $B=8$. The number $J$ of S-boxes for each round is given by $\frac{K}{8}$. One can see that, with a small number $r$ of rounds, the BER is generally smaller for larger $K$, because it takes more rounds for the case of large $K$ to spread the errors over the entire bits. However, for larger number of rounds (e.g., $r \geq 4$), the BER is essentially 0.5 regardless of the size $K$ of the SPN.

\begin{figure}[h]
\begin{center}
\includegraphics[width=0.7\columnwidth]{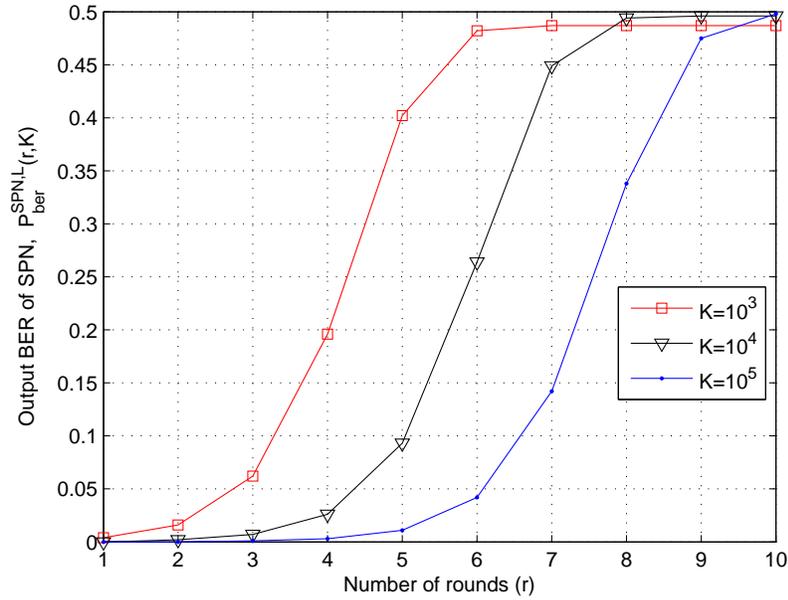}
\caption{BER at the output of SPN composed of $8 \times 8$ {\it practical} S-boxes that are adopted for AES \cite[Fig. 3.8]{Stinson_book}, when only one input bit is in error out of total $K$ input bits. The number $J$ of S-boxes for each round is given by $\frac{K}{8}$. The BER is numerically obtained by simulation.}
\label{fig:SPN_BER_sim}
\end{center}
\end{figure}

Above analysis and numerical results are based on the ideal S-boxes satisfying SAC. We now evaluate the BER of an actual SPN composed of real S-boxes. In this paper, as an example, we use the actual $8 \times 8$ S-boxes adopted for AES \cite[Fig. 3.8]{Stinson_book}, which is known to have good avalanche property \cite{H_Shi}. For the case of single input bit error, Fig. \ref{fig:SPN_BER_sim} shows the output BER obtained by simulations.
One can see that, by increasing the number $r$ of rounds, it is possible to make the output BER close to 0.5. This means that the input error can be effectively amplified by actual SPNs.

\section{Secure Transmission in BER Sense with Finite Blocklength}
In this section, by combining the strong converse and cryptographic confusion and diffusion, a transmission scheme that is secure in the BER sense is proposed. Then the rate margins, security gains, and power optimization are discussed.

\subsection{Combining Strong Converse and Cryptographic Confusion and Diffusion}

When $C_\eve < R< C_\bob$, by increasing blocklength $n$, it is possible to make Bob's block error probability arbitrarily small and Eve's block error probability arbitrarily large. Ensuring small block error probability at Bob means reliable communication. However, ensuring high block error probability at Eve does not necessarily mean that the transmission is secure, because a block error event simply means that there is at least a single bit error in the block. As a simple example, one may consider the case where only a single bit within a codeword is always in error whenever the codeword is decoded. In this case,  the block error probability is one; however, all other bits except the one are decoded by Eve, which means the communication is never secure. In order to address this issue, a method to induce high BER at Eve is discussed.

\begin{figure}[h]
\begin{center}
\includegraphics[width=1.0\columnwidth]{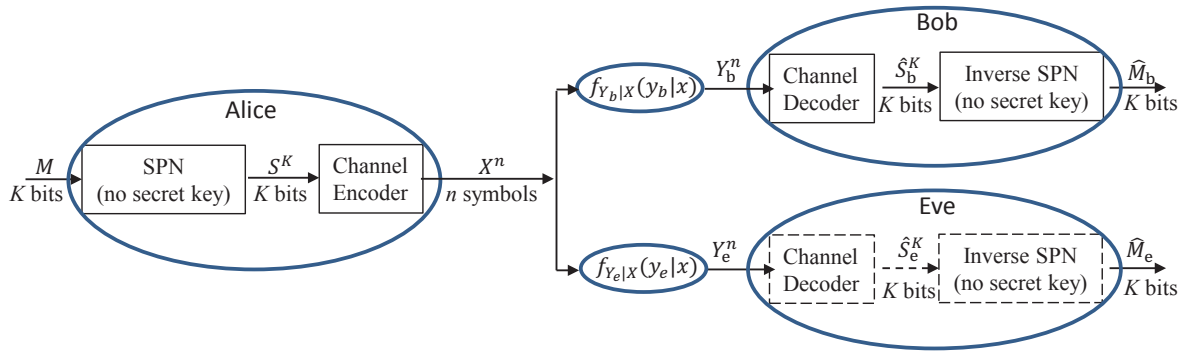}
\caption{Block diagram of the proposed scheme.}
\label{fig:block_diagram}
\end{center}
\end{figure}

The block diagram of the proposed scheme is presented in Fig. \ref{fig:block_diagram}. Using an SPN, Alice encrypts message $M$ of length $K$ bits into bit sequence $S^K$ of the same length, which is then encoded into a codeword $X^n$ of length $n$ symbols. Bob performs the inverse processing: he decodes the received codeword $Y_\bob^n$ into bit sequence $\hat S_\bob^K$, which is then decrypted into $\hat M_\bob$ by the inverse SPN. On the other hand, in principle, Eve can design her receiver as she wants, no matter what it is. In this paper, Eve's receiver structure is assumed to be the same as Bob's, which appears to be a reasonable assumption because otherwise it seems even more difficult for her to estimate $M$. Eve decodes the received codeword $Y_\eve^n$ into bit sequence $\hat S_\eve^K$, which is then decrypted into $\hat M_\eve$ by the inverse SPN.

At the receiver side (Bob or Eve), if no block decoding error occurs at the channel decoder, there is no bit error at the input of the inverse SPN, and thus, no output bit errors. On the other hand, when block decoding error occurs at the channel decoder, there is at least one  bit error at the input of the inverse SPN and the input error(s) will be amplified by the inverse SPN. The BER $P_\ber(R|\calC)$ for a code $\calC$ at the output of the inverse SPN is given by
\beqa  P_\ber(R|\calC) &=& \left. P_{\ber}^{\rm SPN}(r,K) \right|_{\rm block ~ error} \times P_\err(R|\calC) \eeqa where $P_\err(R|\calC)$ denotes the block error probability at the output of the decoder and $\left. P_\ber^{\rm SPN}(r,K) \right|_{\rm block ~ error}$ is the BER at the output of the inverse SPN given a block error. In order to (analytically or numerically) compute the BER $\left. P_\ber^{\rm SPN}(r,K) \right|_{\rm block ~ error}$, the initial error distribution $q_{W_0}(w_0)$ must be determined from the condition that there was a block error, which means that there was at least a single bit error at the input of the inverse SPN. However, the exact number of bit errors within a block is random and the exact distribution of the number of bit errors is unknown. Furthermore, the exact block error probabilities, $P_\err(R|\calC)$, for Bob and Eve are unknown. In the following, therefore, we consider their bounds: for Bob, an upper-bound of the ensemble average $\mathbb{E}[P_\err(R|\calC)]$ is used; and for Eve, a lower-bound of $P_\err(R|\calC)$ is used.

For Bob, using ensemble average block error probability upper-bound $P_\err^\bob(R) \leq P_\err^{\bob, \up}(R, \breve \rho,\breve q(x))$ in (\ref{eq:upper_bound_Bob}) and noting that $\left. P_{\ber}^{\rm SPN}(r,K) \right|_{\rm block ~ error}$ is upper-bounded by 0.5,\footnote{Although 0.5 is a trivial BER upper-bound, it is actually tight in our case because the output BER $\left. P_{\ber}^{\rm SPN}(r,K) \right|_{\rm block ~ error}$  of SPN given a block error is (very) close to 0.5 as long as $r$ is large enough, e.g., $r \geq 10$, as shown in Figs. \ref{fig:SPN_BER_ana} and \ref{fig:SPN_BER_sim}.} the ensemble average BER of Bob is upper-bounded as follows:
\beqa  P_\ber^\bob (R)  &\leq & 0.5 P_\err^{\bob,\up}(R,\breve \rho,\breve q(x)) \\ &=:& P_\ber^{\bob, \up}(R,\breve \rho,\breve q(x)). \eeqa
For Eve, using $P_\err^\eve(R|\calC) \geq P_\err^{\eve, \low}(R,\breve \rho', \breve q'(x)), \forall \calC$ in (\ref{eq:lower_bound_Eve}) and $P_{\ber}^{\rm SPN}(r,K) \geq P_{\ber}^{\rm SPN,\low}(r,K)$, the BER is lower-bounded as follows:
\beqa  P_\ber^\eve(R|\calC)   &\geq & P_{\ber}^{\rm SPN,\low} (r,K) P_\err^{\eve,\low}(R, \breve \rho', \breve q'(x)), ~~~ \forall \calC \\ &=:& P_\ber^{\eve, \low}(R, \breve \rho', \breve q'(x)) \eeqa where  $P_{\ber}^{\rm SPN,\low}(r,K)$ is given by \beq P_{\ber}^{\rm SPN,\low}(r,K) = \left. P_{\ber}^{\rm SPN}(r,K)\right|_{\rm only ~ one ~input ~ bit ~ error}. \eeq That is, $P_{\ber}^{\rm SPN,\low}(r,K)$ denotes the BER at the output of the inverse SPN when there is only a single input bit error (rather than at least one input bit error), and $P_{\ber}^{\rm SPN,\low}(r,K)$ can be obtained by analysis or simulation as in Section II.C.

In general, the optimal distribution $\breve q(x)$ making the upper bound tightest and the distribution $\breve q'(x)$ making the lower bound valid for any $\calC$ are not necessarily the same, i.e., $\breve q(x) \neq \breve q'(x)$. For symmetric DMCs, however, they are the same and given by equi-probable distributions.

{\it Lemma 1:} For symmetric DMCs including BSC, BEC, and binary input (BI)-AWGN, we have
\beq \breve q(x) =   \breve q'(x) = q_{\rm equ}(x) \eeq
where $q_{\rm equ}(x)$ is the equi-probable distribution.

{\it Proof:} See Appendix A. \hfill $\Box$

In our scheme, two bounds are imposed at the same time given a single transmitter (Alice). Therefore, it is important to ensure the existence of such code satisfying both bounds. That is, it must be ensured that at least a code exists for which Bob's BER (not Bob's ensemble average BER) is upper-bounded by $P_\ber^{\bob,\up}(R, \breve \rho, \breve q(x))$ and Eve's BER is lower-bounded by $P_\ber^{\eve,\low}(R, \breve \rho', \breve q'(x))$ at the same time. Such existence is shown in the following.

{\it Lemma 2:} When $C_\eve < R < C_\bob$ and a priori probabilities are the same, \beqa \exists \calC {\rm ~~such ~that~~}  P_\ber^\bob (R|\calC)  \leq P_\ber^{\bob, \up}(R,\breve \rho,\breve q(x)) {\rm ~~and~~} P_\ber^\eve(R|\calC)  \geq P_\ber^{\eve, \low}(R, \breve \rho', \breve q'(x)). \eeqa

{\it Proof:} When $R < C_\bob$, there exists {\it at least one} code for which Bob's BER is upper-bounded by $P_\ber^{\bob,\up}(R, \breve \rho, \breve q(x))$. Furthermore, when $R>C_\eve$ and a priori probabilities are the same, Eve's BER for {\it any} code is lower-bounded by $P_\ber^{\eve,\low}(R, \breve \rho', \breve q'(x))$. Therefore, there must exist a code satisfying both. \hfill $\Box$

For the asymptotic case of infinite blocklength, we have $\lim_{n \rightarrow \infty} P_\err^{\bob,\up}(R, \breve \rho, \breve q(x))=0$ and $\lim_{n \rightarrow \infty}P_\err^{\eve,\low}(R, \breve \rho', \breve q'(x)) =1$ when $C_\eve<R<C_\bob$. Thus, Bob's BER upper-bound and Eve's BER lower-bound are asymptotically given by
\beqa \lim_{n\rightarrow \infty}   P_\ber^{\bob,\up}(R, \breve \rho, \breve q(x))  & = & 0, ~~~~~ R<C_\bob \\ \lim_{n\rightarrow \infty}  P_\ber^{\eve,\low}(R, \breve \rho', \breve q'(x))  & = & P_{\ber}^{\rm SPN,\low}(r,K), ~~~~~ R>C_\eve. \eeqa Recall that $P_{\ber}^{\rm SPN,\low}(r,K)$ can be made very close to 0.5 by increasing the number $r$ of rounds, as demonstrated in Figs. \ref{fig:SPN_BER_ana} and \ref{fig:SPN_BER_sim}.

\subsection{Rate Upper and Lower Bounds for Finite Blocklength}

In practice, the blocklength $n$ is finite, and thus, it is not possible to achieve $P_\ber^{\bob,\up}(R,\breve \rho, \breve q(x))\rightarrow 0$ when $R<C_\bob$. In this paper, therefore, Bob's BER upper-bound is constrained  to be smaller than a BER threshold, $0<\calP_\ber^{\bob,\Th} \leq  0.5$, as follows:
\beqa  P_\ber^\bob(R) \leq P_\ber^{\bob, \up}(R,\breve \rho, \breve q(x))  \leq  \calP_\ber^{\bob, \Th}.  \label{eq:reliability_cond_init} \eeqa
This condition will be referred to as the reliability condition.
To adjust $\calP_\ber^{\bob,\Th}$, it is possible to use a block error probability threshold $0<{\cal P}_\err^{\bob,\Th} \leq 1$, which is related to $\calP_\ber^{\bob,\Th}$ as follows: \beq \calP_\ber^{\bob,\Th} = 0.5 {\cal P}_\err^{\bob,\Th}. \eeq
For high reliability, ${\cal P}_\err^{\bob,\Th}$
should be set small (e.g., $10^{-6}$).  Similar to Bob's case, with finite blocklength $n$, it is not possible to achieve $P_\ber^{\eve, \low}(R,\breve \rho', \breve q'(x)) \rightarrow P_\ber^{\rm SPN,\low}(r,K)$ for Eve when $R>C_\eve$. Therefore, Eve's BER lower-bound is constrained to be larger than a BER threshold, $\calP_\ber^{\eve,\Th}$ with $0 \leq \calP_\ber^{\eve,\Th} < P_{\ber}^{\rm SPN,\low}(r,K)$ as follows:\beqa   P_\ber^\eve (R|
\calC) \geq P_\ber^{\eve, \low}(R,\breve \rho', \breve q'(x)) \geq \calP_\ber^{\eve, \Th}, ~~~ \forall \calC. \label{eq:security_cond_init} \eeqa This condition will be referred to as the security condition. To adjust $\calP_\ber^{\eve,\Th}$, it is possible to use a block error probability threshold  $0 \leq {\cal P}_\err^{\eve,\Th}<1$, which is related to $\calP_\ber^{\eve,\Th}$ as follows: \beq \calP_\ber^{\eve,\Th} = P_{\ber}^{\rm SPN, \low}(r,K) {\cal P}_\err^{\eve,\Th}. \eeq For high security,  ${\cal P}_\err^{\eve,\Th}$ should be set large (e.g., $0.999999$).

When the reliability condition is imposed, the highest possible rate is lower than $C_\bob$. Also, when the security condition is imposed, the lowest possible rate is higher than $C_\eve$. In the following, the rate differences are defined.

{\it Definition 1:} The rate margin from above is defined by $\Delta R_\bob:=C_\bob - R_{\sup}$ and the rate margin from below is defined by $\Delta R_\eve := R_{\inf} - C_\eve$, where the highest allowable transmission rate $R_{\sup}$ and the lowest allowable transmission rate $R_{\inf}$ are determined by
\beqa R_{\sup } & = & \sup_{0 \leq R < C_\bob} R  {\rm ~~~ subject ~ to} ~  P_\ber^{\bob, \up}(R,\breve \rho, \breve q(x))  \leq    \calP_\ber^{\bob, \Th}
\label{eq:R_max_optimization} \\ R_{\inf } &=&  \inf_{R > C_\eve } R  {\rm ~~~ subject ~ to} ~  P_\ber^{\eve, \low}(R,\breve \rho', \breve q'(x))  \geq    \calP_\ber^{\eve, \Th}.   \label{eq:R_min_optimization} \eeqa
\hfill $\Box$

In the following theorem, $\Delta R_\bob$ and $\Delta R_\eve$ are analyzed.

{\it Theorem 1}: For ${\cal P}_\err^{\bob,\Th} =1$, we have  $\Delta R_\bob = 0$. For $0< {\cal P}_\err^{\bob,\Th} <1$, we have
\beqa \Delta R_\bob &=&  -\frac{1}{n \breve \rho} \ln {\cal P}_\err^{\bob, \Th} + C_\bob - \frac{1}{\breve \rho}  E_0^\bob( \breve \rho, \breve q(x)) \\ & \geq &  -\frac{1}{n \breve \rho} \ln {\cal P}_\err^{\bob, \Th} \label{eq:R_b_bound} \\ &>&  0  \eeqa where optimal $\breve \rho$ is determined by $\breve \rho = \arg \max_{0 < \rho \leq 1} \left\{ E_0^\bob(\rho, \breve q(x))-\rho R_{\sup} \right\}$. For ${\cal P}_\err^{\eve,\Th} =0$, we have $\Delta R_\eve = 0$. For $0<{\cal P}_\err^{\eve,\Th} < 1$, we have
\beqa   \Delta R_\eve &=& \frac{1}{n \breve \rho'} \ln \left(1-{\cal P}_\err^{\eve,\Th}\right) + \frac{1}{ \breve \rho'}  E_0^\eve(\breve \rho', \breve q'(x))-C_\eve \\ & \geq  & \frac{1}{n \breve \rho'} \ln \left(1-{\cal P}_{\err}^{\eve,\Th} \right) \label{eq:R_e_bound} \\  & >&  0 \eeqa where optimal $\breve \rho'$ is determined by $\breve \rho' = \arg \max_{-1 < \rho' <0} \left\{ E_0^\eve(\rho', \breve q'(x))-\rho' R_{\inf} \right\}$. As $n \rightarrow \infty$, both rate margins tend to zero: $\Delta R_\bob \rightarrow 0$ and $\Delta R_\eve \rightarrow 0$.

{\it Proof:} See Appendix B.
\hfill $\Box$

From Theorem 1, one can see that, when $0< {\cal P}_\err^{\bob,\Th} <1$, the rate marge from above $\Delta R_\bob$ is always positive, inversely proportional to $n$ and $\breve \rho$, and logarithmically inversely proportional to $\calP_\err^{\bob, \Th}$.  Thus, to reduce $\Delta R_\bob$, it appears that increasing the blocklength would be more effective  than increasing $\calP_\err^{\bob, \Th}$. A similar observation can be made for the rate margin from below $\Delta R_\eve$.

Let $\Delta \calR = R_{\sup} - R_{\inf}$ denote the rate interval in which the actual transmit rate $R$ can be chosen. When $\Delta \calR > 0$, it is possible for Alice to transmit data {\it reliably} and {\it securely} satisfying (\ref{eq:reliability_cond_init}) and (\ref{eq:security_cond_init}). However, if $\Delta \calR < 0$, it is not possible to choose a rate $R$ satisfying both conditions at the same time, and the data transmission is suspended. Letting $\Delta \calC = C_\bob - C_\eve$ denote the capacity interval, the difference between capacity and rate intervals is given by $\Delta \calC - \Delta \calR = \Delta R_\bob + \Delta R_\eve >0$. For the case of fading channels, the intervals are random variables and we have $\Pr(\Delta \calC<0) < \Pr(\Delta \calR<0)$, meaning that the data suspension probability increases with shorter blocklength and stronger reliability/security conditions.

{\it Remark 1:} Ideally, the rate margins should have been defined using the constraints $P_\ber^\bob(R) \leq \calP_\ber^{\bob,\Th}$ and $P_\ber^\eve(R|\calC) \geq \calP_\ber^{\eve,\Th}$, rather from $P_\ber^{\bob, \up}(R,\breve \rho, \breve q(x))  \leq    \calP_\ber^{\bob, \Th}$ and \break $P_\ber^{\eve, \low}(R,\breve \rho', \breve q'(x))  \geq    \calP_\ber^{\eve, \Th}$ as in Theorem 1. Thus, the results of Theorem 1 can be interpreted as follows: There exists {\it at lease one} code whose rate margins from above and below are {\it not} larger than $\Delta R_\bob$ and $\Delta R_\eve$, respectively.


\subsection{Security Gap}
For some specific codes over BI-AWGN channels, the security gap was defined as the difference between Bob's received signal to noise ratio (SNR) required to ensure Bob's BER smaller than a threshold and Eve's received SNR required to ensure Eve's BER larger than a threshold \cite{d_klinc,M_Baldi}. In general, the smaller the security gap, the suitable and more efficient the code for secure communications based on the BER security measure. By simulating specifically designed punctured-LDPC codes for BI-AWGN channels, the authors of \cite{d_klinc} numerically obtained the security gap for their own codes. Similarly, in \cite{M_Baldi}, the security gap was numerically obtained by simulating some specific BCH and LDPC codes combined with scrambling/descrambling for BI-AWGN channels. In this subsection, a fundamental limit of the security gap for any code with finite blocklength is studied for our proposed secure communications of combining strong converse and error amplification.

Consider the unconstrained Gaussian channel, where the received signals at Bob and Eve are given by \beqa Y_{\bob,i}&=& X_i + \eta_{\bob,i}, \ \ i=1,\cdots, n \\ Y_{\eve,i} &=&  X_i + \eta_{\eve,i}, \ \ i=1,\cdots, n \eeqa where $\eta_{\bob,i} \sim {\cal CN}(0, \sigb)$ and $\eta_{\eve,i} \sim {\cal CN}(0, \sige)$ represent AWGNs at Bob and Eve, respectively. The transmitted signal $X_i$ is normalized such that $\mathbb{E}[|X_i|^2]=1$. Then the SNRs at Bob and Eve, respectively, are given by $\gamma_\bob = \frac{\mathbb{E}[|X_i|^2]}{\sigb}=\frac{1}{\sigb}$ and $\gamma_\eve = \frac{\mathbb{E}[|X_i|^2]}{\sige}= \frac{1}{\sige}$. We now define the security gap as follows.

{\it Definition 2:} For AWGN channels, the security gap $\Delta S$ is defined by \beqa && \Delta S  :=  10 \log_{10} \frac{\gamma_{\bob}^{\inf}}{\gamma_{\eve}^{\sup}} \label{eq:security_gap_GI_AWGN} \eeqa where the lowest SNR $\gamma_\bob^{\inf}$ for Bob and the highest SNR $\gamma_\eve^{\sup}$ for Eve are determined by \beqa \gamma_{\bob}^{\inf} &=& \inf_{\gamma_\bob > \gamma_0} \gamma_\bob {\rm ~~~subject ~ to ~~~}  P_\ber^{\bob, \up}(R,\breve \rho,\breve q(x),\gamma_\bob) \leq \calP_\ber^{\bob,\Th} \label{eq:gamma_G_min} \\ \gamma_{\eve}^{\sup} &=& \sup_{0 \leq \gamma_\eve < \gamma_0} \gamma_\eve  {\rm ~~~subject ~ to ~~~} P_\ber^{\eve, \low}(R,\breve \rho',\breve q'(x), \gamma_\eve) \geq \calP_\ber^{\eve,\Th} . \label{eq:gamma_G_max}  \eeqa
In the above equations, $\gamma_0 = C_{\rm AWGN}^{-1}(R)$, where $C_{\rm AWGN}(\gamma)=\ln(1+\gamma)$ denotes the capacity of AWGN channels.
\hfill $\Box$

In order to determine Bob's tightest ensemble average BER upper-bound and Eve's valid BER lower bound for any code, the optimal input distributions $\breve q(x)$ and $\breve q'(x)$ must be first determined by maximizing $E_0^\bob(\rho, q(x), \gamma_\bob)$ and minimizing $E_0^\eve(\rho', q'(x), \gamma_\eve)$, respectively. Such optimizations are generally challenging, because the optimizations should be numerically performed and the optimal distributions depend on $\gamma_\bob$, $\gamma_\eve$, and $R$ (through $\breve \rho$ and $\breve \rho'$). In this subsection, for analytical tractability, we choose the input distributions as ${\cal CN}(0, 1)$, which is denoted by $q_{\cal CN}(x)$.
With $q_{\cal CN}(x)$, the upper-bound of Bob's ensemble average BER and the lower-bound of Eve's BER are given in closed-form as follows: \beqa P_\ber^{\bob,\up}(R, \rho,q_{\cal CN}(x),\gamma_\bob) &=& 0.5 \exp \left( -n  \left\{ E_0^\bob(\rho, q_{\cal CN}(x), \gamma_\bob) - \rho R \right\} \right)\label{eq:Bob_BER_Gau_real} \\ P_\ber^{\eve,\low}(R, \rho',q_{\cal CN}(x),\gamma_\eve) &=& P_\ber^{\rm SPN,\low}(r, K) \cdot \left( 1-\exp \left( -n \left\{ E_0^\eve(\rho', q_{\cal CN}(x), \gamma_\eve) - \rho' R \right\} \right)  \right) \label{eq:Eve_BER_Gau_real} \eeqa where \beqa E_0^\bob(\rho, q_{\cal CN}(x), \gamma_\bob) & =& -\ln \left( 1+ \frac{ \gamma_\bob  }{ 1+\rho  } \right)^{-\rho}, ~~~~ 0 \leq \rho \leq 1 \\ E_0^\eve(\rho', q_{\cal CN}(x), \gamma_\eve) & =& -\ln \left( 1+ \frac{ \gamma_\eve  }{ 1+\rho'  } \right)^{-\rho'}, ~~~~ -1 < \rho' \leq 0. \eeqa

To determine $\Delta S$ for AWGN channels, the highest SNR $\gamma_\bob^{\inf}$ and the lowest SNR $\gamma_\eve^{\sup}$ are first obtained in the following lemma.

{\it Lemma 3:} The solutions to (\ref{eq:gamma_G_min}) and (\ref{eq:gamma_G_max}) with $\breve q(x)=\breve q'(x) =q_{\cal CN}(x)$ are given by  \beqa \gamma_\bob^{\inf} &=& \left\{ \begin{array}{ll} \gamma_0, & {\rm if ~} {\cal P}_\err^{\bob,\Th} =1 \\ g_\bob(\breve \rho), & {\rm if ~} 0< {\cal P}_\err^{\bob,\Th} <1 \end{array} \right. \\ \gamma_\eve^{\sup} &=& \left\{ \begin{array}{ll} \gamma_0, & {\rm if~} {\cal P}_\err^{\eve,\Th} =0 \\ g_\eve(\breve \rho'), & {\rm if ~} 0< {\cal P}_\err^{\eve,\Th} <1 \end{array} \right. \eeqa
where  \beqa g_\bob (\rho) &=&  (1+\rho) \left( \left( {\cal P}_\err^{\bob,\Th} \right)^{-\frac{1}{n \rho}} e^{R} -1 \right)  \label{eq:g_Bob} \\ g_\eve(\rho')&=&  (1+ \rho')  \left(\left(1-{\cal P}_\err^{\eve,\Th}\right)^{-\frac{1}{n \rho'}} e^{R} -1\right).  \label{eq:g_Eve}
\eeqa Optimal $\breve \rho$ and $\breve \rho'$ are determined by \beqa \breve \rho &=&\arg \min_{0 < \rho \leq 1}  g_\bob (\rho) \label{eq:optimization_rho_1} \\ \breve \rho' &=& \arg \max_{-1< \rho' <0}  g_\eve (\rho'). \label{eq:optimization_rho_prime_1}  \eeqa
The optimal solution $\breve \rho$ to (\ref{eq:optimization_rho_1}) always exists for left-open interval $(0,1]$ and $g_\bob (\breve \rho) > \gamma_0$. The optimal solution $\breve \rho'$ to (\ref{eq:optimization_rho_prime_1}) always exists for open interval $(-1,0)$ and $g_\bob (\breve \rho') < \gamma_0$. Also $g_\bob (\breve \rho')$ is positive if and only if the following condition is satisfied:
\beq \left(1-\frac{2}{n} \ln(1-{\cal P}_\err^{\eve,\Th}) \right) \left( 1- {\cal P}_\err^{\eve,\Th} \right)^{\frac{1}{n}} e^R >1 \label{eq:existence_cond_optimization_1}.  \eeq
{\it Proof:} See Appendix C. \hfill $\Box$

From the lemma, we immediately have the following result.

{\it Theorem 2:} When $0< {\cal P}_\err^{\bob,\Th} <1$ and $0< {\cal P}_\err^{\eve,\Th} <1$, the security gap $\Delta S$ with $\breve q(x)=\breve q'(x) =q_{\cal CN}(x)$ is given by
\beqa \Delta S =
10 \log_{10}\frac{(1+\breve \rho) \left( \left( {\cal P}_\err^{\bob,\Th} \right)^{-\frac{1}{n \breve \rho}} e^{R} -1 \right)}{(1+ \breve \rho')  \left(\left(1-{\cal P}_\err^{\eve,\Th}\right)^{-\frac{1}{n \breve \rho'}} e^{R} -1\right)} \label{S_gap_GI_AWGN}  \eeqa where $\breve \rho$ and $\breve \rho'$ are given by (\ref{eq:optimization_rho_1}) and (\ref{eq:optimization_rho_prime_1}), respectively.
\hfill $\Box$

It is not difficult to show $\lim_{n \rightarrow \infty} \Delta S =0$, which one can expect. Also, if one takes a high SNR approximation assuming $\gamma_\bob \gg 1$ and $ \gamma_\eve \gg 1$, it is easier to obtain analytical insights into the security gap.
When $\gamma_\bob \gg 1$ and $\gamma_\eve \gg 1$, the upper-bound of Bob's ensemble average BER and the lower-bound of Eve's BER can be approximated as follows: \beqa P_\ber^{\bob,\up}(R, \rho,q_{\cal CN}(x),\gamma_\bob) &\simeq & 0.5 \exp \left( -n  \left\{  -\ln \left( \frac{ \gamma_\bob  }{ 1+\rho  } \right)^{-\rho} - \rho R \right\} \right), ~~0 \leq \rho \leq 1 \label{eq:Bob_BER_Gau_approx} \\ P_\ber^{\eve,\low}(R, \rho',q_{\cal CN}(x),\gamma_\eve) & \simeq & P_\ber^{\rm SPN,\low}(r, K) \cdot \left( 1-\exp \left( -n \left\{ -\ln \left( \frac{ \gamma_\eve  }{ 1+\rho'  } \right)^{-\rho'} - \rho' R \right\} \right)  \right),   \non \\ && \hspace{7.5cm} -1< \rho' \leq 0. \label{eq:Eve_BER_Gau_approx} \eeqa From the approximate BER bounds, the security gap is obtained as follows:
\beqa \Delta S & \simeq & -\frac{1}{n \breve \rho}10 \log_{10} {\cal P}_\err^{\bob,\Th}+ \frac{1}{n \breve \rho'} 10 \log_{10}\left(1- {\cal P}_\err^{\eve,\Th}\right) + 10 \log_{10} \left(
\frac{1+\breve \rho}{1+\breve \rho'} \right)   \label{eq:security_gap_approx} \eeqa where $0<\breve \rho \leq 1$ and $-1 < \breve \rho' < 0$. From this expression, one can easily see that the security gap is inversely proportional to $n$ and logarithmically inversely proportional to $\calP_\err^{\bob, \Th}$ and $(1-\calP_\err^{\eve, \Th})$.
Note that it is incorrect to interpret (\ref{eq:security_gap_approx}) to mean that, because $R$ does not explicitly appear in (\ref{eq:security_gap_approx}), $\Delta S$ becomes independent of $R$ in high SNR. Since both $\breve \rho$ and $\breve \rho'$ depend on $R$, the security gap $\Delta S$ still depends on $R$ in high SNR.

{\it Remark 2 (Input distribution):}
Gaussian distribution does not necessarily maximize $E_0^\bob(\rho,q(x),\gamma_\bob)$, $0\leq \rho \leq 1$ for all $\gamma_\bob$ and $R < C_\bob$. Thus, the ensemble average BER upper-bound (\ref{eq:Bob_BER_Gau_real}) is not necessarily the tightest one. Nevertheless, the upper-bound is still valid in the sense that there exists a code for which Bob's BER is upper-bounded by (\ref{eq:Bob_BER_Gau_real}). Similarly, Gaussian distribution does not necessarily minimize $E_0^\eve(\rho',q'(x),\gamma_\eve), -1 < \rho' \leq 0$ for all $\gamma_\eve$ and $R>C_\eve$. In this case, the BER lower-bound might not be valid in the sense that the BER of some codes might not be lower-bounded by (\ref{eq:Eve_BER_Gau_real}). Consequently, for the particular code(s) whose BER at Bob is upper-bounded by (\ref{eq:Bob_BER_Gau_real}) for all $\gamma_\bob$ and $R< C_\bob$, the corresponding BER at Eve might not be always larger than (\ref{eq:Eve_BER_Gau_real}) for all $\gamma_\eve$ and $R>C_\eve$. In this sense,  Eve's BER lower-bound of (\ref{eq:Eve_BER_Gau_real}) is optimistic. Nevertheless, using Gaussian input distribution is still useful because it makes the analysis tractable and gives some insights. Furthermore, it satisfies the asymptotic property (\ref{eq:asympt_rho_prime_1}) of the distribution, which makes Eve's block error probability lower-bound valid for all codes, as follows:
\beqa \lim_{\rho' \uparrow 0} \frac{1}{\rho'}  E_0^\eve(\rho',q_{\cal CN}(x),\gamma_\eve)  =  \ln \left( 1+ \gamma_\eve \right)= C_\eve.   \eeqa
This means that Gaussian input distribution makes (\ref{eq:Eve_BER_Gau_real}) valid for any code when $\rho' \uparrow 0$, which is optimal $\rho'$ when $R \rightarrow C_\eve$ from above. Furthermore, Gaussian distribution satisfies the asymptotic property  (\ref{eq:asympt_rho_1}) of the distribution, which makes Bob's ensemble average block error probability upper-bound tightest, as follows:
\beqa  \lim_{\rho \downarrow 0} \frac{1}{\rho}  E_0^\eve(\rho,q_{\cal CN}(x),\gamma_\bob) =  \ln \left( 1+ \gamma_\bob \right)= C_\bob.   \eeqa
This means that Gaussian input distribution makes (\ref{eq:Bob_BER_Gau_real}) tightest when $\rho \downarrow 0$, which is optimal $\rho$ when $R \rightarrow C_\bob$ from below.

{\it Remark 3 ($\texttt{M}$-ary input AWGN): }
For one-dimensional or two-dimensional $\texttt{M}$-ary discrete input AWGN channels with equi-input probabilities, the security gap $\Delta S$ can be obtained by (\ref{eq:security_gap_GI_AWGN}), (\ref{eq:gamma_G_min}), and (\ref{eq:gamma_G_max}) by using $\gamma_0 = C_{\rm MI-AWGN}^{-1}(R)$, where $C_{\rm MI-AWGN}(\gamma)$ denotes the capacity of the $\texttt{M}$-ary input AWGN channel given in \cite[eq. (1.20)]{William}. Also, Bob's ensemble average BER upper-bound and Eve's BER lower-bound can be obtained in a similar way as in the Gaussian input case. As an example, for BI-AWGN, the bounds  are given by
\beqa && P_\ber^{\bob,\up}(R,\rho,q_{\rm equ}(x), \gamma_\bob) = 0.5 \times \non \\ && \exp \left( -n  \left\{  -\ln \left[ \int_{-\infty}^\infty \sqrt{\frac{\gamma_\bob}{2\pi} } \exp \left( -\frac{1}{2} \gamma_\bob(y_{\bob}^2 +1) \right)  \left( \cosh\left( \frac{\gamma_\bob y_\bob}{1 + \rho} \right) \right)^{1+\rho} d y_\bob \right]  - \rho R \right\} \right) \label{eq:Bob_BER_bi_Gau}   \\ && P_\ber^{\eve,\low}(R,\rho',q_{\rm equ}(x),\gamma_\eve) = P_\ber^{\rm SPN,\low}(r,K) \times \non \\ && \left( 1-\exp \left( -n \left\{  -\ln \left[ \int_{-\infty}^\infty \sqrt{\frac{\gamma_\eve}{2\pi} } \exp \left( - \frac{1}{2} \gamma_\eve(y_{\eve}^2 +1) \right)  \left( \cosh\left( \frac{\gamma_\eve y_\eve}{1 + \rho'} \right) \right)^{1+\rho'} d y_\eve \right]  - \rho' R \right\} \right)  \right) \non \\ \label{eq:Eve_BER_bi_Gau} \eeqa
where $0\leq  \rho \leq 1$ and $-1 <\rho' \leq 0$ are optimized to obtain tightest bounds. Unlike the Gaussian input case, it is difficult to analytically obtain the security gap $\Delta S$ for the $\texttt{M}$-ary input case because the BER bounds are not given in closed-form. Thus, $\Delta S$ should be obtained numerically.


{\it Remark 4 (BSC and BEC):} Although the security gap was originally considered only for AWGN channels in the literature, the concept can be extended to other channels such as BSC and BEC. Let $\varepsilon_\bob$ denote the crossover and erasure probabilities for BSC and BEC, respectively, for Bob. Let $\varepsilon_\eve$ denote the crossover and erasure probabilities for BSC and BEC, respectively, for Eve. It is assumed that $0\leq \varepsilon_\bob < \varepsilon_\eve \leq 0.5$ for BSC, and $0\leq \varepsilon_\bob < \varepsilon_\eve \leq 1$ for BEC. Given $R$, the security gap can be defined as the difference between the two probabilities as follows:

{\it Definition 3:} For BSC and BEC, the security gap is defined as follows:
\beqa && \Delta S  :=  \varepsilon_{\eve}^{\inf} - \varepsilon_{\bob}^{\sup} \geq 0 \eeqa where $\varepsilon_\bob^{\sup}$ and $\varepsilon_\eve^{\inf}$ are determined by \beqa \varepsilon_{\bob}^{\sup} &=& \sup_{0 \leq \varepsilon_\bob < \varepsilon_0} \varepsilon_\bob {\rm ~~~subject ~ to ~~~}  P_\ber^{\bob, \up}(R,\breve \rho,q_{\rm equ}(x), \varepsilon_\bob) \leq \calP_\ber^{\bob,\Th} \\ \varepsilon_{\eve}^{\inf} &=& \inf_{\varepsilon_\eve > \varepsilon_0} \varepsilon_\eve  {\rm ~~~subject ~ to ~~~}  P_\ber^{\eve, \low}(R,\breve \rho',q_{\rm equ}(x), \varepsilon_\eve) \geq \calP_\ber^{\eve,\Th}.  \eeqa
In these equations, $\varepsilon_0 = C_{\rm BSC}^{-1}(R)$ for BSC and $\varepsilon_0 = C_{\rm BEC}^{-1}(R)$ for BEC, where $C_{\rm BSC}(\varepsilon)$ and $C_{\rm BEC}(\varepsilon)$ are the capacities of BSC and BEC, respectively.
\hfill $\Box$

For BSC, the BER bounds are given by \beqa && P_\ber^{\bob,\up}(R,\rho, q_{\rm equ}(x),  \varepsilon_\bob) \non \\ &&= 0.5 \exp \left( -n \left\{ -\ln \left[ 2^{-\rho} \left( \varepsilon_\bob^{\frac{1}{1+\rho}}+(1-\varepsilon_\bob)^{\frac{1}{1+\rho}} \right)^{1+\rho} \right]-\rho R \right\} \right) \\ && P_\ber^{\eve,\low}(R,\rho',q_{\rm equ}(x), \varepsilon_\eve) \non \\ &&= P_\ber^{\rm SPN, \low}(r,K)\cdot \left( 1- \exp \left( -n \left\{ -\ln \left[ 2^{-\rho'} \left( \varepsilon_\eve^{\frac{1}{1+\rho'}}+(1-\varepsilon_\eve)^{\frac{1}{1+\rho'}} \right)^{1+\rho'} \right]-\rho' R  \right\} \right) \right). \non \\ \eeqa
For BEC, the BER bounds are given by \beqa  P_\ber^{\bob,\up}(R,\rho, q_{\rm equ}(x),  \varepsilon_\bob)&=& 0.5 \exp \left( -n \left\{ -\ln \left[ 2^{-\rho} (1-\varepsilon_\bob) +\varepsilon_\bob \right] -\rho R \right\} \right) \\  P_\ber^{\eve,\low}(R,\rho',q_{\rm equ}(x), \varepsilon_\eve) &=& P_\ber^{\rm SPN, \low}(r,K)\cdot \left( 1- \exp \left( -n \left\{ -\ln \left[ 2^{-\rho'} (1-\varepsilon_\eve) +\varepsilon_\eve \right] -\rho' R  \right\} \right) \right). \non \\ \eeqa



\subsection{Power Optimization for Gaussian-Input Fading Channels}

In this subsection, the transmit power is optimized for Gaussian-input fading channels. Let $h_\bob$ denote the channel from Alice to Bob and $h_\eve$ the channel from Alice to Eve, where $h_\bob$ and $h_\eve$ are fixed over the duration of a codeword. The received signals at Bob and Eve are given by \beqa Y_{\bob,i}&=&h_\bob X_i + \eta_{\bob,i}, \ \ i=1,\cdots, n \\ Y_{\eve,i} &=& h_\eve X_i + \eta_{\eve,i}, \ \ i=1,\cdots, n \eeqa where $\eta_{\bob,i} \sim {\cal CN}(0, \sigb)$ and $\eta_{\eve,i} \sim {\cal CN}(0, \sige)$. The transmit power $p$ is given by $p=\mathbb{E}[|X_i|^2]$. Let $\Gamma_\bob = \frac{|h_\bob|^2}{\sigb}$ denote Bob's instantaneous channel SNR and $\Gamma_\eve = \frac{|h_\eve|^2}{\sige}$ denote Eve's instantaneous channel SNR. When the input distribution is given by $q_{\cal CN}(x) ={\cal CN}(0, p)$, the upper-bound of Bob's ensemble average BER and the lower-bound of Eve's BER are given in closed-form as follows: \beqa P_\ber^{\bob,\up}(R,\rho,q_{\cal CN}(x), \Gamma_\bob, p) &=& 0.5 \exp \left( -n  \left\{  -\ln \left( 1+ \frac{ p \Gamma_\bob  }{ 1+\rho  } \right)^{-\rho} - \rho R \right\} \right), ~~0 \leq \rho \leq 1 \label{eq:Bob_BER_Gau_comp} \\ P_\ber^{\eve,\low}(R,\rho',q_{\cal CN}(x), \Gamma_\eve, p) &=& P_\ber^{\rm SPN,\low}(r,K) \cdot \left( 1-\exp \left( -n \left\{ -\ln \left( 1+ \frac{ p \Gamma_\eve  }{ 1+\rho'  } \right)^{-\rho'} - \rho' R \right\} \right)  \right),   \non \\ && \hspace{8.5cm} -1< \rho' \leq 0. \label{eq:Eve_BER_Gau_comp} \eeqa

Using these bounds, we first define the reliability, security, and overall outage probabilities as follows:

{\it Definition 4:} The reliability outage is declared whenever $P_\ber^{\bob,\up}(R, \breve \rho, q_{\cal CN}(x), \Gamma_\bob, p) >   \calP_\ber^{\bob,\Th}$, the security outage is declared whenever $P_\ber^{\eve,\low}(R, \breve \rho', q_{\cal CN}(x), \Gamma_\eve, p) <   \calP_\ber^{\eve,\Th}$, and the overall outage is declared whenever $P_\ber^{\bob,\up}(R,\breve \rho,q_{\cal CN}(x),\Gamma_\bob, p) >   \calP_\ber^{\bob,\Th}$  or $P_\ber^{\eve,\low}(R, \breve \rho', q_{\cal CN}(x), \Gamma_\eve, p) <   \calP_\ber^{\eve,\Th}$. The reliability, security, and overall outage probabilities are given by
\beqa
P^{\rm rel}_{\rm out}(R, p) &=& \Pr\left( P_\ber^{\bob,\up}(R, \breve \rho, q_{\cal CN}(x), \Gamma_\bob, p) >   \calP_\ber^{\bob,\Th} \right) \label{eq:rel_out_prob_definition}\\
P^{\rm sec}_{\rm out}(R,p) &=& \Pr\left( P_\ber^{\eve,\low}(R, \breve \rho', q_{\cal CN}(x), \Gamma_\eve, p) <   \calP_\ber^{\eve,\Th} \right)  \label{eq:sec_out_prob_definition} \\
P_{\rm out}^{\rm overall}(R, p) &=&  \Pr\left( P_\ber^{\bob,\up}(R,\breve \rho,q_{\cal CN}(x),\Gamma_\bob, p) >   \calP_\ber^{\bob,\Th}{\rm ~ or ~} P_\ber^{\eve,\low}(R, \breve \rho', q_{\cal CN}(x), \Gamma_\eve, p) <   \calP_\ber^{\eve,\Th} \right) \label{eq:out_prob_definition}. \non \\
\eeqa
\hfill $\Box$

Now, the transmit power is optimized to minimize the reliability outage probability subject to an average power constraint and the security condition for Eve:  \begin{subequations} \beqa && \min_{p(\Gamma_\bob, \Gamma_\eve)} P^{\rm rel}_{\rm out}(R, p(\Gamma_\bob,\Gamma_\eve)) \\ && {\rm subject ~ to} ~  p(\Gamma_\bob, \Gamma_\eve) \geq 0  \\ && \hspace{1.9cm} \mathbb{E}[p(\Gamma_\bob, \Gamma_\eve)] \leq p_\av   \\ && \hspace{1.9cm}  P_\ber^{\eve,\low}(R,\breve \rho' ,q_{\cal CN}(x), \Gamma_\eve, p(\Gamma_\bob,\Gamma_\eve)) \geq \calP_\ber^{\eve,\Th}  \eeqa \label{eq:prob_1_opt_1}
\end{subequations} where the transmit power $p(\Gamma_\bob, \Gamma_\eve)$ is denoted as an explicit function of $\Gamma_\bob$ and $\Gamma_\eve$.

When $\calP_\ber^{\bob,\Th}=0.5$ or ${\cal P}_\err^{\bob,\Th}=1$, the reliability outage probability is always zero. Also, when $\calP_\ber^{\eve,\Th}=0$ or ${\cal P}_\err^{\eve,\Th}=0$, the security constraint is degenerate. Therefore, focusing on $0< {\cal P}_\err^{\bob,\Th} <1$ and $0< {\cal P}_\err^{\eve,\Th} <1$, the optimal solution is derived in the following.

{\it Theorem 3:} For $0< {\cal P}_\err^{\bob,\Th} <1$ and $0< {\cal P}_\err^{\eve,\Th} <1$, the optimal solution to (\ref{eq:prob_1_opt_1}) is given by
\beq p_{\opt} (\Gamma_\bob, \Gamma_\eve) = \left\{ \begin{array}{ll} p_{\min}(\Gamma_\bob, \breve \rho), & {\rm if ~} p_{\min}(\Gamma_\bob,\breve \rho) \leq p_{\max}(\Gamma_\eve,\breve \rho')  {\rm ~and ~} p_{\min}(\Gamma_\bob,\breve \rho) \leq z_\opt  \\
0, & {\rm if ~} p_{\min}(\Gamma_\bob,\breve \rho) > p_{\max}(\Gamma_\eve,\breve \rho')  {\rm ~~or ~~} p_{\min}(\Gamma_\bob,\breve \rho) > z_\opt  \end{array} \right. \label{eq:solution_optimization_1} \eeq
where \beqa p_{\min} (\Gamma_\bob,\rho) &=&   \Gamma_\bob^{-1} g_\bob (\rho)  \label{eq:p_min_Bob} \\ p_{\max}(\Gamma_\eve,\rho')&=&  \Gamma_\eve^{-1} g_\eve (\rho')  \label{eq:p_max_Eve}
\\  z_\opt &=& \max \{z: z \geq 0, \mathbb{E}[p_{\opt}(\Gamma_\bob, \Gamma_\eve)] \leq p_\av \}. \eeqa In the above equations, $g_\bob(\rho)$, $g_\eve(\rho')$, $\breve \rho'$, and $\breve \rho'$ are respectively given by (\ref{eq:g_Bob}), (\ref{eq:g_Eve}), (\ref{eq:optimization_rho_1}), and (\ref{eq:optimization_rho_prime_1}).

{\it Proof:} See the Appendix D. \hfill $\Box$

The optimal power $p_{\opt}(\Gamma_\bob, \Gamma_\eve)$ derived in Theorem 3 can be intuitively explained as follows. Firstly, in order to avoid any reliability outage at Bob, at least certain amount of transmit power should be used. Given $\Gamma_\bob$, power $p_{\min} (\Gamma_\bob,  \rho)$ is the minimum instantaneous power required to satisfy Bob's reliability condition $P_\ber^{\bob,\up}(R,\rho,q_{\cal CN}(x),\Gamma_\bob,p) \leq  \calP_\ber^{\bob,\Th}$. However, when we consider Eve, too much transmit power leads to  weak security, because with higher power she can more easily decode the codeword.  In order to enhance security for Eve and eventually to avoid any security outage at Eve, less transmit power should be used to ensure lower SNR at Eve. Given $\Gamma_\eve$, power $p_{\max} (\Gamma_\eve, \rho')$ is the maximum instantaneous allowable power to satisfy the security condition $P_\ber^{\eve,\low}(R,\rho',q_{\cal CN}(x),\Gamma_\eve,p) \geq \calP_\ber^{\eve,\Th}$. Overall, any transmit power in the interval $[p_{\min} (\Gamma_\eve, \rho), p_{\max} (\Gamma_\bob, \rho')]$ satisfies both reliability and security conditions. With the average power constraint, however, the transmit power must be set to a minimum possible level by  $p_{\opt}(\Gamma_\bob, \Gamma_\eve)=p_{\min} (\Gamma_\eve, \rho)$, to minimize the reliability outage probability by most efficiently utilizing the power on average.

Secondly, the case of $p_{\opt}(\Gamma_\bob, \Gamma_\eve)=0$ in Theorem 3 can be explained as follows. When $p_{\min} (\Gamma_\bob, \rho)$ is greater than $p_{\max} (\Gamma_\bob, \rho')$, it is not possible to satisfy both the reliability and security requirements at the same time, and thus, the data transmission must be suspended i.e., $p_{\opt}(\Gamma_\bob,\Gamma_\eve)=0$. Furthermore, due to the average power constraint, the transmission is suspended by setting $p_{\opt}(\Gamma_\bob,\Gamma_\eve)=0$ whenever the minimum power required is too large, i.e., $p_{\min}(\Gamma_\bob,  \rho) > z$. Let $P_{\rm sus}(R, p(\Gamma_\bob, \Gamma_\eve))$ denote the data transmission suspension probability given by
\beqa P_{\rm sus}(R, p(\Gamma_\bob, \Gamma_\eve)) = \Pr( p(\Gamma_\bob,\Gamma_\eve)=0). \eeqa
Then $z$ is maximized under the average power constraint in order to minimize the suspension probability, which is a necessary condition for reliability outage probability minimization because reliability outage occurs whenever $p(\Gamma_\bob,\Gamma_\eve)=0$. The condition of $p_{\min}(\Gamma_\bob,  \rho) > z_\opt$ can be rewritten as \beq \Gamma_\bob < \frac{g_\bob( \rho)}{z_\opt}. \eeq  This means that, for efficient power consumption, the data transmission must be suspended when Bob's instantaneous channel SNR $\Gamma_\bob$ is worse than a threshold.

Finally, the reason why optimal $\breve \rho$ and $\breve \rho'$ in Theorem 3 are obtained by (\ref{eq:optimization_rho_1}) and (\ref{eq:optimization_rho_prime_1}), respectively, can be explained as follows. In order to minimize $P_{\rm sus}(R, p(\Gamma_\bob, \Gamma_\eve))$ or $\Pr( p(\Gamma_\bob,\Gamma_\eve)=0)$, the value of $\rho$ must be optimized to minimize $p_{\min} (\Gamma_\bob,  \rho)$, which is equivalent to the optimization in (\ref{eq:optimization_rho_1}). Also, $\rho'$ must be optimized to maximize $p_{\max} (\Gamma_\eve, \rho')$, which is equivalent to the optimization in (\ref{eq:optimization_rho_prime_1}). The computational complexities required for these optimization are not high because each of $\breve \rho$ and $\breve \rho'$ can be individually obtained by one-dimensional searching.

From Theorem 3, we also have the following result.

{\it Corollary 1:} With the optimal power $p_{\opt}(\Gamma_\bob, \Gamma_\eve)$ of Theorem 3, we have
\beqa P^{\rm rel}_{\rm out}(R,p_{\opt}(\Gamma_\bob, \Gamma_\eve) ) & = & P_{\rm out}^{\rm overall}(R, p_{\opt}(\Gamma_\bob, \Gamma_\eve)) =P_{\rm sus}(R, p_{\opt}(\Gamma_\bob, \Gamma_\eve))   \\P^{\rm sec}_{\rm out}(R, p_{\opt}(\Gamma_\bob, \Gamma_\eve)) &=& 0. \eeqa

{\it Proof:} First, when $p_{\opt}(\Gamma_\bob, \Gamma_\eve)=0$, we have $\Pr \left( P_\ber^{\eve,\low}\left(R, \breve \rho',q_{\cal CN}(x), \Gamma_\eve, 0 \right) <   \calP_\ber^{\eve,\Th} \right)=0$, because $P_\ber^{\eve,\low}(R, \breve \rho',q_{\cal CN}(x), \Gamma_\eve, 0 )=0.5$ with probability one. Second, when $p_{\opt}(\Gamma_\bob, \Gamma_\eve)=p_{\min}(\Gamma_\bob,\breve \rho)$, we have
$\Pr \left( P_\ber^{\eve,\low}(R, \breve \rho',q_{\cal CN}(x),\Gamma_\eve, p_{\min}(\Gamma_\bob,\breve \rho) )   <    \calP_\ber^{\eve,\Th} \right)=0$, because  $P_\ber^{\eve,\low}(R, \breve \rho', \break q_{\cal CN}(x), \Gamma_\eve, p_{\min}(\Gamma_\bob,\breve \rho) ) =\calP_\ber^{\eve,\Th}$ with probability one as shown in Appendix D. It follows from the total probability theorem that  $P_{\rm out}^{\rm sec}(R, p_{\opt}(\Gamma_\bob, \Gamma_\eve))=0$. Given this, it is straightforward to show $P^{\rm rel}_{\rm out}(R, p_{\opt}(\Gamma_\bob, \Gamma_\eve))  =  P_{\rm out}^{\rm overall}(R, p_{\opt}(\Gamma_\bob, \Gamma_\eve))=P_{\rm sus}(R, p_{\opt}(\Gamma_\bob, \Gamma_\eve)) $. \hfill $\Box$



\section{Numerical Results}
\begin{figure}[h]
\begin{center}
\includegraphics[width=0.7\columnwidth]{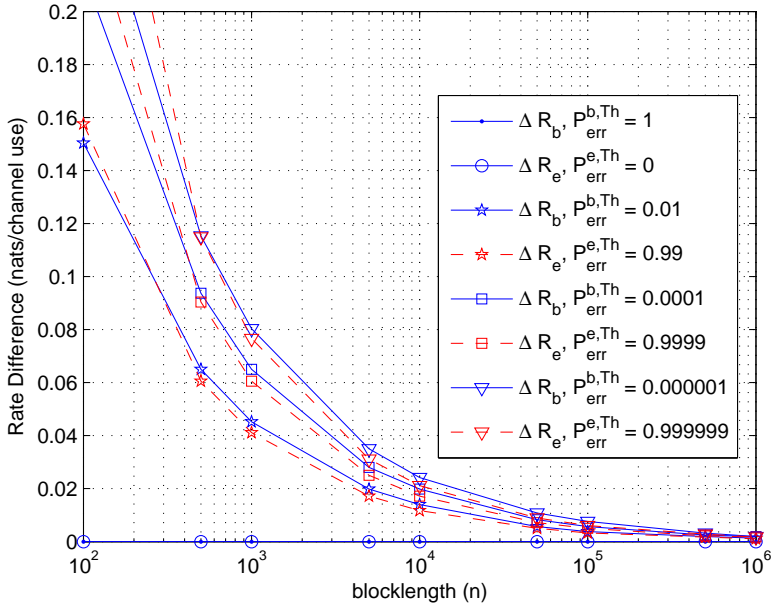}
\caption{Rate differences $\Delta R_\bob$ and $\Delta R_\eve$ for BSC with $\varepsilon_\bob = 0.01$ and $\varepsilon_\eve = 0.3$.}
\label{fig:max_min_rate_BSC}
\end{center}
\end{figure}

\begin{figure}[h]
\begin{center}
\includegraphics[width=0.7\columnwidth]{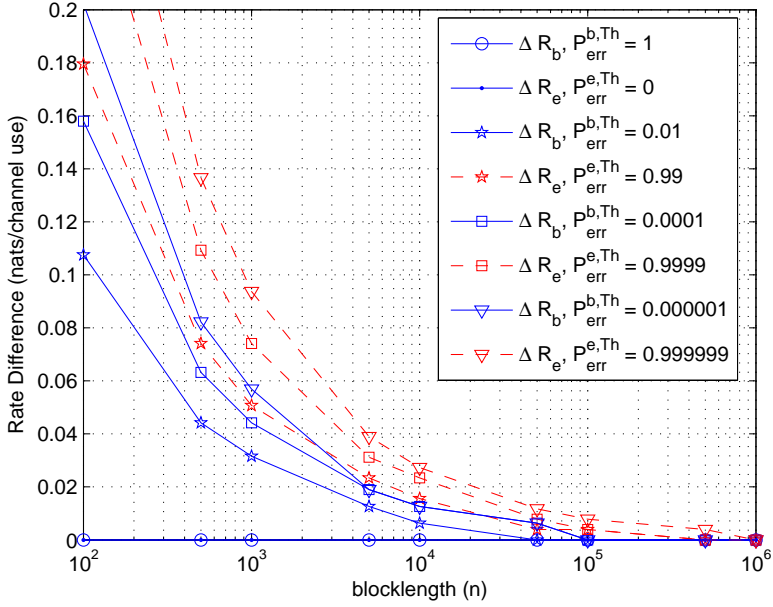}
\caption{Rate differences $\Delta R_\bob$ and $\Delta R_\eve$ for BI-AWGN with $\gamma_\bob = 6$ dB and $\gamma_\eve = -2$ dB.}
\label{fig:max_min_rate_BI_AWGN}
\end{center}
\end{figure}

In this section, we present numerical results for the proposed secure communication method. First, we present the numerical results of rate margins $\Delta R_\bob$ and $\Delta R_\eve$ obtained in Theorem 1. Different reliability and security requirements are tested by considering ${\cal P}_\err^{\bob,\Th} \in \{1, 0.01, 0.0001, 0.000001\}$ and ${\cal P}_\err^{\eve,\Th} \in \{0, 0.99, 0.9999, 0.999999\}$. Recall that Bob's BER upper-bound threshold is given by $\calP_\ber^{\bob,\Th} = 0.5{\cal P}_\err^{\bob,\Th}$ and Eve's BER lower-bound threshold is given by $\calP_\ber^{\eve,\Th} = P_\ber^{\rm SPN, \low}(r,K) {\cal P}_\err^{\eve,\Th}$. Fig. \ref{fig:max_min_rate_BSC} shows $\Delta R_\bob$ and $\Delta R_\eve$ for BSC with $\varepsilon_\bob = 0.01$ and $\varepsilon_\eve = 0.3$ for different blocklengths $10^2 \leq n  \leq 10^6$. Also, Fig. \ref{fig:max_min_rate_BI_AWGN} shows $\Delta R_\bob$ and $\Delta R_\eve$ for BI-AWGN with $\gamma_\bob = 6$ dB and $\gamma_\eve = -2$ dB. One can see that as the blocklength $n$ increases, $\Delta R_\bob$ and $\Delta R_\eve$ approach zero as expected in Theorem 1. With weaker reliability and security requirements (i.e., larger ${\cal P}_\err^{\bob,\Th}$ and smaller ${\cal P}_\err^{\eve,\Th}$), the rate margins decrease.

\begin{figure}[h]
\begin{center}
\includegraphics[width=0.7\columnwidth]{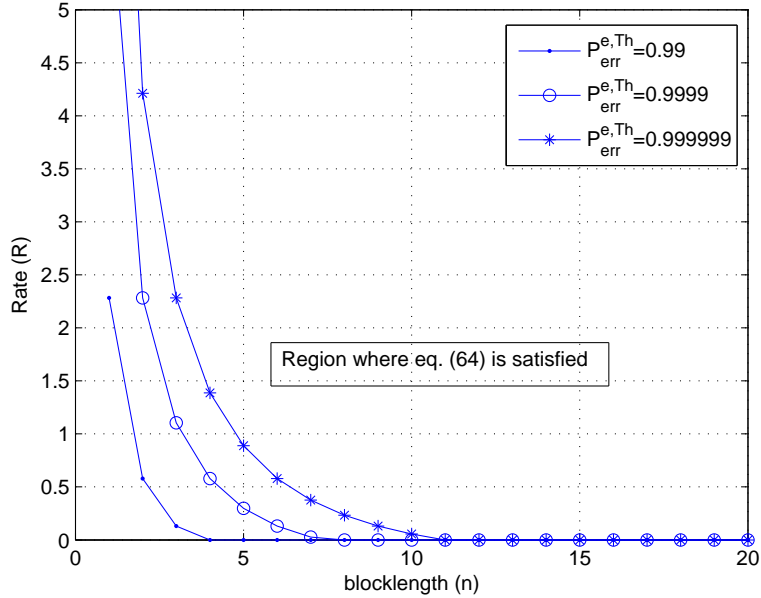}
\caption{Region where eq. (\ref{eq:existence_cond_optimization_1}) is satisfied.}
\label{fig:feasibility}
\end{center}
\end{figure}

\begin{figure}[h]
\begin{center}
\includegraphics[width=0.7\columnwidth]{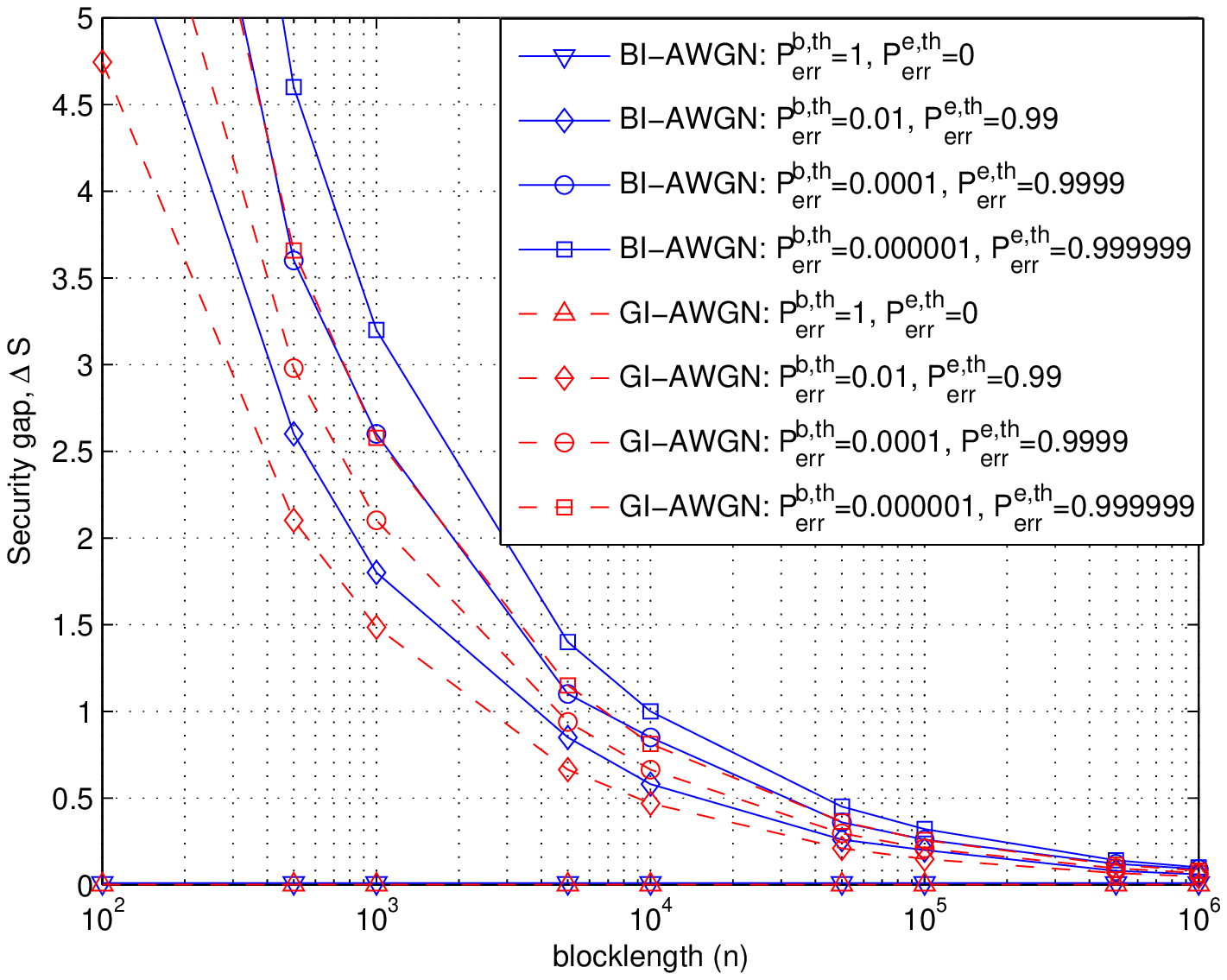}
\caption{Security gap $\Delta S$ for BI-AWGN with $R=0.5$ (nats/one-dimensional-channel use) and for GI-AWGN with $R=1$ (nats/two-dimensional-channel use).}
\label{fig:S_gap_BI_GI}
\end{center}
\end{figure}

Second, we present the numerical results for the security gap $\Delta S$ obtained in Theorem 2. Fig. \ref{fig:feasibility} shows the region where the condition of (\ref{eq:existence_cond_optimization_1}) is satisfied. It can be easily seen that the condition is satisfied for all practical cases, e.g., for all $R$ with $n>10$. Fig. \ref{fig:S_gap_BI_GI} shows the security gaps for BI-AWGN with $R=0.5$ (nats/one-dimensional-channel use) and Gaussian-input (GI) AWGN with $R=1$ (nats/two-dimensional-channel use). It can be seen that, for the same reliability and security conditions, Gaussian input gives smaller security gap than binary input.

\begin{figure}[h]
\begin{center}
\includegraphics[width=0.7\columnwidth]{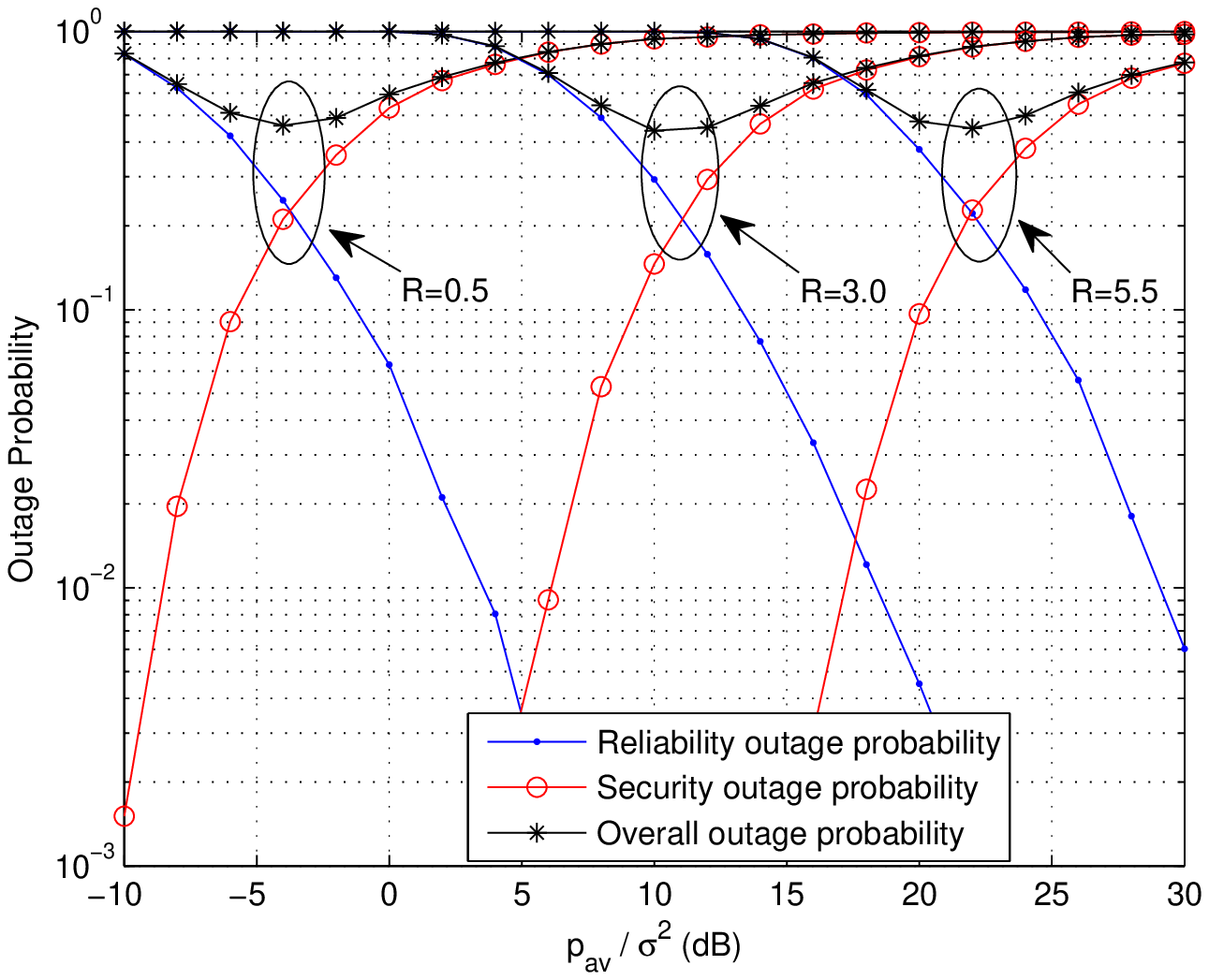}
\caption{Reliability outage probability $P_{\rm out}^{\rm rel}(R,p_\av)$ of (\ref{eq:rel_out_prob_definition}), security outage probability $P_{\rm out}^{\rm sec}(R,p_\av)$ of (\ref{eq:sec_out_prob_definition}), and overall outage probability $P_{\rm out}^{\rm overall}(R,p_\av)$ of (\ref{eq:out_prob_definition}) with constant transmit power with $p=p_\av$. ${\cal P}_\err^{\bob,\Th}=0.0001$ and ${\cal P}_\err^{\eve,\Th}=0.9999$. $\mathbb{E}[|h_\bob|^2]=2$ and $\mathbb{E}[|h_\eve|^2]=1$. Blocklength $n$ is $10^5$.}
\label{fig:rel_minimization_equ_pow}
\end{center}
\end{figure}

We now consider fading channels, where $|h_\bob|$ and $|h_\eve|$ are modeled by Rayleigh random variables with $\mathbb{E}[|h_\bob|^2]=2$ and $\mathbb{E}[|h_\eve|^2]=1$ and we set $\sigma_\bob^2 = \sigma_\eve^2 = \sigma^2$. Generating the channels $10^5$ times, we numerically obtain the reliability outage probability $P_{\rm out}^{\rm rel}(R,p)$ of (\ref{eq:rel_out_prob_definition}), the security outage probability $P_{\rm out}^{\rm sec}(R,p)$ of (\ref{eq:sec_out_prob_definition}), and the overall outage probability $P_{\rm out}^{\rm overall}(R,p)$ of (\ref{eq:out_prob_definition}).  Note that when $|h_\bob| < |h_\eve|$ (i.e., $C_\bob < C_\eve$), it is never possible to avoid both reliability and security outages at the same time no matter which rate $R$ and transmit power $p$ are used. In the following, therefore, we evaluate the outage probabilities only when $|h_\bob| > |h_\eve|$. The reliability and security conditions are set to ${\cal P}_\err^{\bob,\Th}=0.0001$ and ${\cal P}_\err^{\eve,\Th}=0.9999$. The blocklength $n$ is set to $10^5$. We first consider the case of equal (or constant) transmit power, where the transmit power $p$ is set to $p_\av$. Fig. \ref{fig:rel_minimization_equ_pow} shows the outage probabilities with the equal transmit power  for different rates $R \in \{0.5, 3.0, 5.5\}$ (nats/two-dimensional-channel use). Given rate $R$, as $p_{\rm av}/\sigma^2$ increases, the reliability outage probability decreases, whereas the security outage probability increases as can be expected. Consequently, the overall outage probability always remains high (e.g., larger than say 0.4), because both reliability and security outage probabilities cannot be decreased at the same time.

\begin{figure}[h]
\begin{center}
\includegraphics[width=0.7\columnwidth]{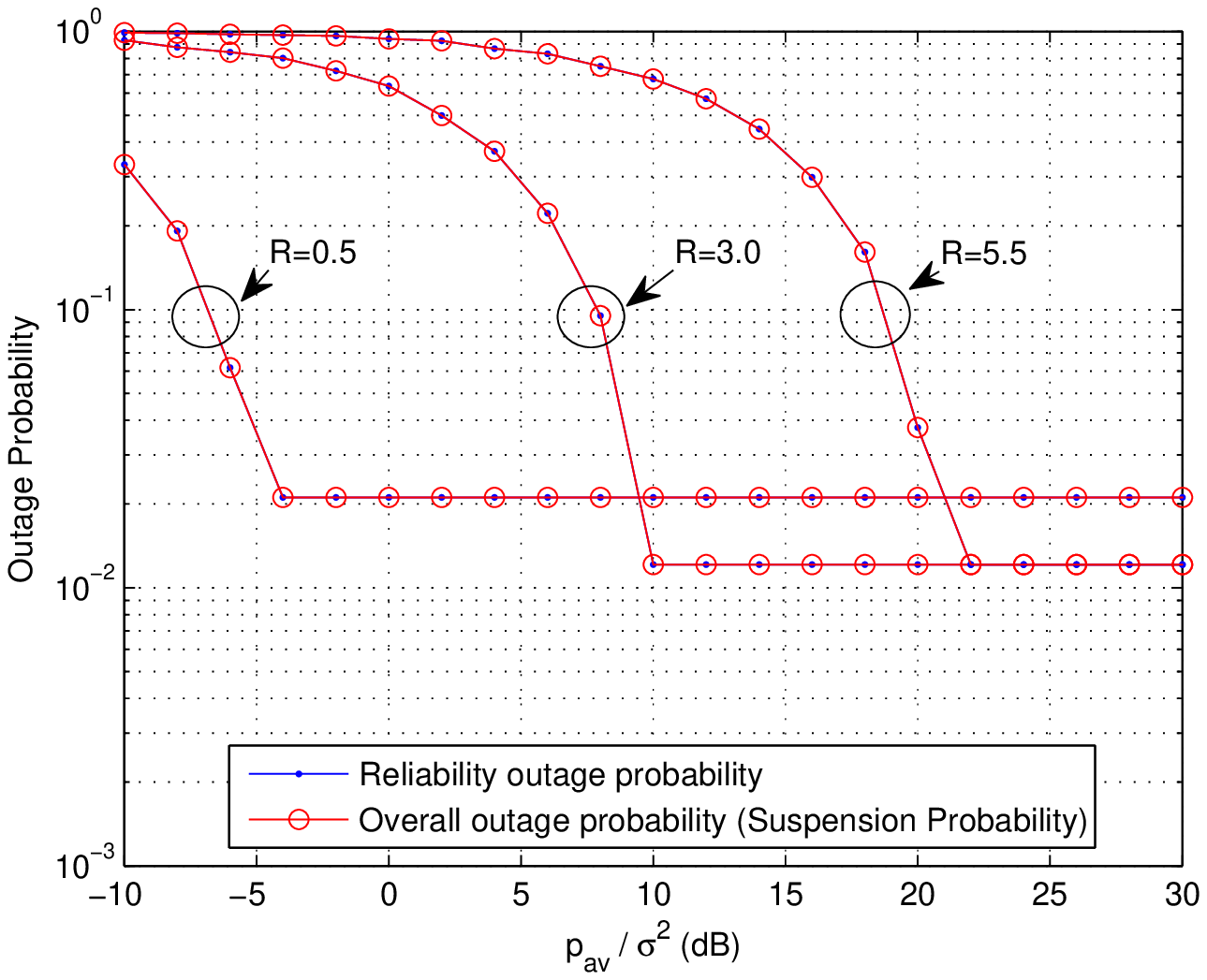}
\caption{Reliability outage probability $P_{\rm out}^{\rm rel}(R,p_\opt(\Gamma_\bob,\Gamma_\eve))$ of (\ref{eq:rel_out_prob_definition}), security outage probability $P_{\rm out}^{\rm sec}(R,p_\opt(\Gamma_\bob,\Gamma_\eve))$ of (\ref{eq:sec_out_prob_definition}), and overall outage probability $P_{\rm out}^{\rm overall}(R,p_\opt(\Gamma_\bob,\Gamma_\eve))$ of (\ref{eq:out_prob_definition}) with optimal power allocation in Theorem 3. ${\cal P}_\err^{\bob,\Th}=0.0001$ and ${\cal P}_\err^{\eve,\Th}=0.9999$. $\mathbb{E}[|h_\bob|^2]=2$ and $\mathbb{E}[|h_\eve|^2]=1$. Blocklength $n$ is $10^5$.}
\label{fig:rel_minimization_opt_pow}
\end{center}
\end{figure}

\begin{figure}[h]
\begin{center}
\includegraphics[width=0.7\columnwidth]{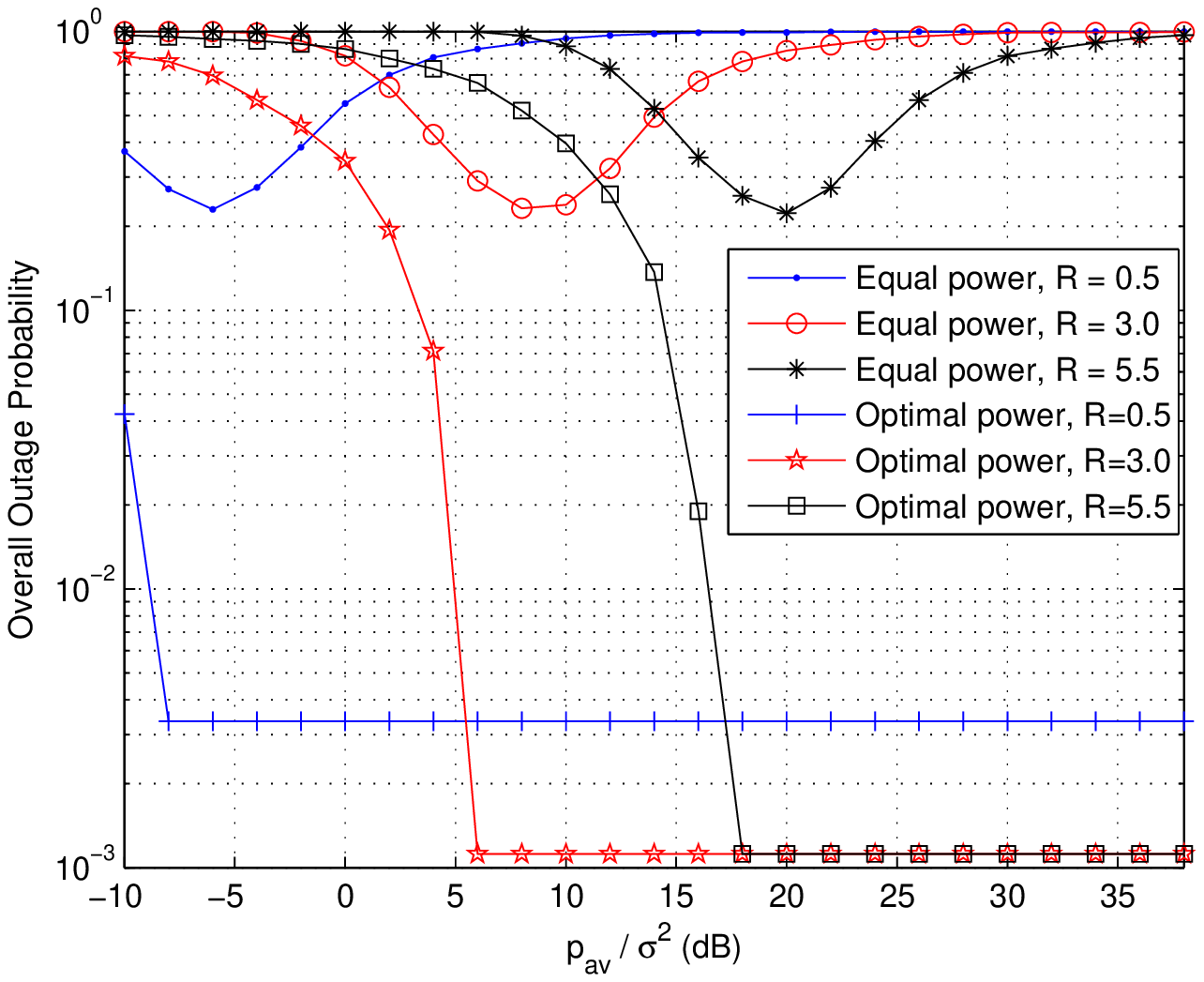}
\caption{Overall outage probabilities of equal transmit power and optimal transmit power.  ${\cal P}_\err^{\bob,\Th}=0.0001$ and ${\cal P}_\err^{\eve,\Th}=0.9999$. $\mathbb{E}[|h_\bob|^2]=10$ and $\mathbb{E}[|h_\eve|^2]=1$. Blocklength $n$ is $10^5$.}
\label{fig:rel_minimization_equ_opt_compare}
\end{center}
\end{figure}

For the proposed optimal power allocation of Theorem 3, Fig. \ref{fig:rel_minimization_opt_pow} shows the outage probabilities with the same system parameters of Fig. \ref{fig:rel_minimization_equ_pow}. The obtained security outage probability of the optimal power allocation is exactly zero, which cannot be plotted in the figure of log-scale outage probability. The reliability probability, the overall outage probability, and the suspension probability are the same, as expected from Corollary 1. As $p_{\rm av}/\sigma^2$ increases, the overall outage probability (or suspension probability) decreases and then flattens.\footnote{This is in sharp contrast to the case of conventional power optimization without a security condition, in which the outage probability (or suspension probability) decreases indefinitely with $p_\av/\sigma^2$ \cite{caire}.} However, the achieved lowest overall outage probability of the optimal power allocation is much lower than that of the constant transmit power case. The reason why there is an error floor for the overall outage probability or suspension probability is as follows. The suspension probability is given by $\Pr(p_{\opt}(\Gamma_\bob,\Gamma_\eve)=0)$. From Theorem 3, the suspension probability can be decreased only by reducing the probabilities of $p_{\min}(\Gamma_\bob,\breve \rho) > p_{\max}(\Gamma_\eve,\breve \rho')$ and $p_{\min}(\Gamma_\bob,\breve \rho) > z_\opt$. By increasing $p_{\rm av}$, it is possible to reduce the probability of  $p_{\min}(\Gamma_\bob,\breve \rho) > z_\opt$. However, it is not possible to reduce the probability of $p_{\min}(\Gamma_\bob,\breve \rho) > p_{\max}(\Gamma_\eve,\breve \rho')$ by increasing power $p_{\rm av}$. In order to reduce the probability of $p_{\min}(\Gamma_\bob,\breve \rho) > p_{\max}(\Gamma_\eve,\breve \rho')$, Bob's channel must be made even better (i.e., larger $\Gamma_\bob$) or Eve's channel must be made even worse (i.e., smaller $\Gamma_\eve$). Only in this case, the suspension outage probability is further decreased. Fig.\ \ref{fig:rel_minimization_equ_opt_compare} shows the overall outage probabilities for the equal power and optimal power allocations for such channel scenario where Bob's channel is much better than Eve's channel: $\mathbb{E}[|h_\bob|^2]=10$ and $\mathbb{E}[|h_\eve|^2]=1$. In this channel scenario, the floors of the overall outage probabilities are lower for both equal and optimal power allocations. But, the performance gap between the two power allocations is still significant.

\section{Conclusions}

In this paper, a secure data transmission method has been studied, where the security measure was given in terms of the BER at the eavesdropper. To realize such secure communication, information-theoretic strong converse and cryptographic error amplification have been combined. For finite blocklengths, the maximum and minimum allowable transmission rates and the security gap have been analyzed for any block codes over DMCs. It has been observed that increasing the blocklength is very effective to reduce the rate loss and the security gap. For fading channels, the transmission power has been optimized. It has been found that simply increasing the transmission power does not decrease the reliability outage probability indefinitely. The error floor of the reliability outage probability depends on the channel quality difference between Bob and Eve.

\clearpage
\newpage

\section*{Appendix A \\ Proof of Lemma 1}
\renewcommand\theequation{A.\arabic{equation}}
\setcounter{equation}{0}

For symmetric DMCs, it is well-known that $E_0^\bob(\rho, q(x)), 0\leq \rho \leq 1,$ is maximized by equi-probable distribution $q_{\rm equ}(x)$  \cite[Theorem 7.2]{jelinek}. In the following, therefore, we will only show that $E_0^\eve(\rho', q'(x)), -1 < \rho' \leq 0,$ is also minimized by $q_{\rm equ}(x)$.


The proof of this appendix is only for finite input and output alphabet sizes. But, the approach holds for well-behaved channels with infinite alphabet sizes. For $x \in \{ a_1,\cdots,a_Q\}$, let us define $\alpha(y_\eve, \bfq)$ as follows \beqa \alpha(y_\eve, \bfq) &=& \sum_x q(x) f_{Y_\eve|X}(y_\eve|x)^{1/(1+\rho')}, ~~~~~ -1 < \rho' \leq 0 \\ &=& \sum_{k=1}^Q q_k f_{Y_\eve|X}(y_\eve|x)^{1/(1+\rho')}, ~~~~~ -1 < \rho' \leq 0 \eeqa where $\bfq=(q_1,\cdots,q_Q) =(q(x=a_1), \cdots, q(x=a_Q))$. Because $\alpha(y_\eve, \bfq)$ is linear in $\bfq$ and the function $\alpha^{1+\rho'}$ is concave in $\alpha$, $\alpha(y, \bfq)^{1+\rho'}$ must be concave in $\bfq$. Letting $F(\rho', \bfq) = \exp(-E_0^\eve(\rho',\bfq))=\sum_{y_\eve} \alpha(y_\eve,\bfq)$, the function $F(\rho', \bfq)$ is  concave, because it is the sum of concave functions. Then $F(\rho',\bfq)$ has a minimum for some $\bfq^0$.

Following \cite[Theorem 4.4.1]{gallager}, the necessary and sufficient conditions that $F(\rho',\bfq)$ is minimized at $\bfq^0$ are \beqa \left. \frac{\partial F(\rho',\bfq)}{\partial q_k} \right|_{\bfq = \bfq^0} & \leq & \lambda, ~~~ k=1,2,\cdots,Q; -1 <\rho' \leq 0 \eeqa where $\lambda$ is a constant and the equality holds whenever $q_k \neq 0$ (i.e., $q_k > 0$). Using \cite[Theorem 5.6.5]{gallager}, the necessary and sufficient conditions on $\bfq$ which maximize $F(\rho', \bfq)$, equivalently, minimize $E_0^\eve(\rho', \bfq)$, are \beq \sum_{y_\eve} f_{Y_\eve|X}(y_\eve|x)^{1/(1+\rho')} \alpha(y_\eve, \bfq)^{\rho'} \leq \sum_{y_\eve} \alpha(y_\eve, \bfq)^{1+\rho'}, ~~~ -1 < \rho' \leq 0  \eeq where equality holds for which $q_k >0$. Finally, for symmetric DMCs, the equi-probable distribution $q_k = \frac{1}{Q}$ satisfies the following condition \cite[Theorem 7.2]{jelinek}: \beq \sum_{y_\eve} f_{Y_\eve|X}(y_\eve|x)^{1/(1+\rho')} \alpha(y_\eve, \bfq)^{\rho'} = \sum_{y_\eve} \alpha(y_\eve, \bfq)^{1+\rho'}, ~~~ -1 < \rho' \leq 0.  \eeq
This complete the proof.

\section*{Appendix B \\Proof of Theorem 1}
\renewcommand\theequation{B.\arabic{equation}}
\setcounter{equation}{0}

For $R_{\sup}$, the optimization problem is \beqa  \sup_{0 \leq R < C_\bob} R  {\rm ~~~ subject ~ to} ~  \min_{0 \leq \rho \leq 1} P_\err^{\bob, \up}(R,\rho, \breve q(x))  \leq   {\cal P}_\err^{\bob,\Th} \eeqa where $P_\err^{\bob, \up}(R,\rho, \breve q(x))=2 P_\ber^{\bob, \up}(R,\rho, \breve q(x))$.
First, we consider the case of ${\cal P}_\err^{\bob,\Th}=1$. In this case, the constraint is always satisfied. Thus, $R_{\sup} = \sup_{0 \leq R < C_\bob}R = C_\bob$. Second, we consider the case of $0< {\cal P}_\err^{\bob,\Th}< 1$. For $\rho=0$, we have $P_\err^{\bob, \up}(R,\rho, \breve q(x))=1$. However, we know that $\min_{0 \leq \rho \leq 1} P_\err^{\bob, \up}(R,\rho, \breve q(x)) < 1$ because $\max_{0 \leq \rho \leq 1} \{ E_0^\bob(\rho, \breve q(x)) -\rho R\} >0$ for $R< C_\bob$. Thus, the optimal $\breve \rho$ must be in $0< \breve \rho \leq 1$. That is, we have
$\breve \rho (R) = \arg \min_{0 < \rho \leq 1} P_\err^{\bob, \up}(R,\rho, \breve q(x)) = \arg \max_{0< \rho\leq 1} \left\{ E_0^\bob(\rho, \breve q(x))-\rho R \right\}$, where $\left\{ E_0^\bob(\rho, \breve q(x))-\rho R \right\}$ is convex in $\rho \in (0,1]$ for $R < C_\bob$ \cite[Proof of Theorem 5.6.3]{gallager}. Because $P_\err^{\bob, \up}(R,\rho, \breve q(x))$ is a monotonically decreasing function of $R$, the constraint must be satisfied with equality to maximize $R$: $P_\err^{\bob, \up}(R_{\sup},\rho,\breve q(x))={\cal P}_\err^{\bob,\Th}$. Thus, we have
\beqa R_{\sup} & = &  \frac{1}{n  \breve \rho(R_{\sup})} \ln {\cal P}_\err^{\bob,\Th}  + \frac{1}{\breve \rho(R_{\sup})}  E_0^\bob( \breve \rho(R_{\sup}), \breve q(x))  \\ & \stackrel{(a)}{\leq} & \frac{1}{n \breve \rho(R_{\sup})} \ln {\cal P}_\err^{\bob,\Th} +  \frac{1}{\breve \rho(R_{\sup})} \max_{q(x)} E_0^\bob( \breve\rho(R_{\sup}), q(x))  \label{eq:R_max_first_eq} \\ & \stackrel{(b)}{\leq} & \frac{1}{n \breve \rho(R_{\sup})} \ln {\cal P}_\err^{\bob,\Th} +   \max_{q(x)}\left. \frac{\partial}{\partial \rho} E_0^\bob(\rho, q(x)) \right|_{\rho=0} \label{eq:R_max_second_eq} \\ & = & \frac{1}{n \breve \rho(R_{\sup})} \ln {\cal P}_\err^{\bob,\Th} +   \max_{q(x)}I_\bob(q(x))    \\ & = & \frac{1}{n \breve \rho(R_{\sup})} \ln {\cal P}_\err^{\bob,\Th} + C_\bob \\ & \stackrel{(c)}{<} & C_\bob \eeqa
where $(b)$ is due to \cite[eq.(34)]{Arimoto} and $(c)$ is valid for any $0<{\cal P}_\err^{\bob,\Th} < 1$. When $n \rightarrow \infty$, we have $R_{\sup} \rightarrow C_\bob$ from below, because the constraint becomes always satisfied by $\min_{0 \leq \rho \leq 1} P_\err^{\bob, \up}(R,\rho, \breve q(x)) \rightarrow 0$ for any $R< C_\bob$ as $n \rightarrow \infty$. Also, when $n \rightarrow \infty$, we have $\Delta R_\bob \rightarrow 0$.

For $R_{\inf}$, the optimization problem is \beqa  \inf_{R > C_\eve} R  {\rm ~~~ subject ~ to} ~  \max_{-1 < \rho' \leq 0} P_\err^{\eve, \low}(R,\rho', \breve q'(x))  \geq  {\cal P}_\err^{\eve,\Th} \eeqa where $P_\err^{\eve, \low}(R,\rho', \breve q'(x))=P_\ber^{\eve, \low}(R,\rho', \breve q'(x))/{\cal P}_\ber^{\rm SPN,\low}(r,K)$. We first consider the case of ${\cal P}_\err^{\eve,\Th}=0$. In this case, the constraint is always satisfied. Thus, $R_{\inf} = \inf_{R > C_\eve}R = C_\eve$. Second, consider the case of $0< {\cal P}_\err^{\eve,\Th}< 1$. For $\rho'=0$, we have $P_\err^{\eve, \low}(R,\rho', \breve q'(x))=0$. However, we know that $\max_{-1 < \rho' \leq 0} P_\err^{\eve, \low}(R,\rho', \breve q'(x)) >0$ because $\max_{-1 < \rho' \leq 0} \{ E_0^\eve(\rho', \breve q'(x)) -\rho' R\} >0$ for $R> C_\eve$. Thus, the optimal $\breve \rho'$ must be in $-1< \breve \rho' <0$.
That is, we have $\breve \rho' (R) = \arg \max_{-1 < \rho <0} P_\err^{\eve, \low}(R,\rho', \breve q'(x)) = \arg \max_{-1<\rho' < 0 }  \left\{ E_0^\eve(\rho', \breve q'(x))-\rho' R \right\}$,
where $\left\{ E_0^\eve(\rho', \breve q'(x))-\rho' R \right\}$ is convex in $\rho' \in (-1,0)$ for $R> C_\eve$ \cite{Arimoto}, \cite[Lemma 3.2.1]{viterbi}. Because  $P_\err^{\eve, \low}(R, \rho', \breve q'(x))$ is a monotonically decreasing function of $R$, the constraint must be satisfied with equality to minimize $R$:
$\max_{-1 < \rho' \leq 0} P_\err^{\eve, \low}(R_{\inf},\rho', \breve q'(x)) = {\cal P}_\err^{\eve,\Th}$. Thus, we have \beqa R_{\inf} & = & \frac{1}{n \breve \rho'(R_{\inf})} \ln \left(1-{\cal P}_\err^{\eve,\Th}\right) + \frac{1}{ \breve \rho'(R_{\inf})}  E_0^\eve(\breve \rho'(R_{\inf}), \breve q'(x))\\ &=&  \frac{1}{n  \rho'(R_{\inf})} \ln \left(1-{\cal P}_\err^{\eve,\Th}\right) + \frac{1}{ \breve \rho'(R_{\inf})} \min_{q(x)}  E_0^\eve(\breve \rho'(R_{\inf}), q(x))   \\ &\stackrel{(d)}{\geq} & \frac{1}{n \breve \rho'(R_{\inf})} \ln \left(1-{\cal P}_\err^{\eve,\Th}\right) +  \max_{q(x)}   \left. \frac{\partial}{\partial \rho'} E_0^\eve(\rho', q(x)) \right|_{\rho'=0}   \\&=& \frac{1}{n \breve \rho'(R_{\inf})} \ln \left(1-{\cal P}_\err^{\eve,\Th}\right) +  \max_{q(x)}   I_\eve(q(x)) \label{eq:R_min_ineq} \\ & = & \frac{1}{n \breve \rho'(R_{\inf})} \ln \left(1-{\cal P}_\err^{\eve,\Th}\right) + C_\eve \\ & \stackrel{(e)}{> } & C_\eve \eeqa
where $(d)$ is due to \cite[eq.(37)]{Arimoto} and $(e)$ is valid for any $0 < {\cal P}_\err^{\eve,\Th} <1$. When $n \rightarrow \infty$, we have $R_{\inf} \rightarrow C_\eve$ from above, because the constraint becomes always satisfied by $\min_{-1 \leq \rho'  \leq 0} P_\err^{\eve, \low}(R,\rho', \breve q'(x)) \rightarrow 1$ for any $R > C_\eve$ as $n \rightarrow \infty$. Thus, when $n \rightarrow \infty$, we have $\Delta R_\eve \rightarrow 0$.


\section*{Appendix C\\Proof of Theorem 2}
\renewcommand\theequation{C.\arabic{equation}}
\setcounter{equation}{0}

For $\gamma_\bob^{\inf}$, the optimization problem is \beqa  \inf_{\gamma_\bob > \gamma_0 } \gamma_\bob {\rm ~~~subject ~ to ~~~} \min_{0 \leq \rho \leq 1} P_\err^{\bob, \up}(R,\rho,q_{\cal CN}(x),\gamma_\bob) \leq  {\cal P}_\err^{\bob,\Th} \eeqa where $P_\err^{\bob, \up}(R,\rho,q_{\cal CN}(x),\gamma_\bob)=2P_\ber^{\bob, \up}(R,\rho,q_{\cal CN}(x),\gamma_\bob)$. First, we consider the case of ${\cal P}_\err^{\bob,\Th}=1$. In this case, the constraint is always satisfied. Thus, $\gamma_\bob^{\inf} = \inf_{\gamma_\bob > \gamma_0 } \gamma_\bob = \gamma_0 $. Second, we consider the case of $0< {\cal P}_\err^{\bob,\Th}< 1$. As shown in Appendix B, the interval for optimizing $\rho$ can be restricted to $0< \rho \leq 1$. For this interval, the constraint can be rewritten as
\beqa \gamma_\bob & \geq &   \min_{0< \rho \leq 1} (1+ \rho) \left( \left( {\cal P}_\err^{\bob,\Th} \right)^{-\frac{1}{n  \rho}} e^{R} -1 \right)= \min_{0< \rho \leq 1} g_\bob(\rho).   \eeqa Thus, $\gamma_\bob^{\inf}=\inf_{\gamma_\bob \geq \gamma_0 } \gamma_\bob = g_\bob(\breve \rho)$, where optimal $\breve \rho$ is given by $\breve \rho = \arg \min_{0< \rho \leq 1} g_\bob(\rho)$. Finally, we show the existence of $\breve \rho$ for left-open interval $(0, 1]$. For all $0 < \rho \leq 1$, it is straightforward to show
\beqa
g_\bob(\rho) & > &  \gamma_0  \geq 0   \\
\lim_{\rho \rightarrow 0+} \frac{\partial}{\partial \rho}g_\bob(\rho) &=& -\infty
\\ \lim_{\rho \rightarrow 1} \frac{\partial}{\partial \rho}g_\bob(\rho) &=&  u_\bob e^R-1 > 0 \\ \frac{\partial^2}{\partial \rho^2}g_\bob(\rho) & > & 0  \eeqa where $ u_\bob = \left( {\cal P}_\err^{\bob,\Th}\right)^{-\frac{1}{n}}  > 1$. Therefore, the optimal $\rho$ minimizing $g_\bob(\rho)$ must not be $\rho \rightarrow 0+$, and a solution exists in $(0, 1]$.

For $\gamma_\eve^{\sup}$, the optimization problem is \beqa  \sup_{0\leq \gamma_\eve < \gamma_0 } \gamma_\eve {\rm ~~~subject ~ to ~~~} \max_{-1 < \rho' \leq 0} P_\err^{\eve, \low}(R,\rho',q_{\cal CN}(x),\gamma_\eve) \geq {\cal P}_\err^{\eve,\Th} \eeqa where $P_\err^{\eve, \low}(R,\rho',q_{\cal CN}(x),\gamma_\eve)=P_\ber^{\eve, \low}(R,\rho',q_{\cal CN}(x),\gamma_\eve)/{\cal P}_\ber^{\rm SPN, \low}(r,K)$.
First, we consider the case of ${\cal P}_\err^{\eve,\Th}=0$. In this case, the constraint is always satisfied. Thus, $\gamma_\eve^{\sup} = \sup_{0 \leq \gamma_\eve < \gamma_0 } \gamma_\eve = \gamma_0 $. Second, we consider the case of $0< {\cal P}_\err^{\eve,\Th}< 1$. As shown in Appendix B, the interval for optimizing $\rho'$ can be restricted to $-1< \rho' <0$. For this interval, the constraint can be rewritten as
\beqa \gamma_\eve & \leq & \max_{-1 < \rho' <0}  (1+  \rho')  \left(\left(1-{\cal P}_\err^{\eve,\Th}\right)^{-\frac{1}{n  \rho'}} e^{R} -1\right)=\max_{-1 < \rho' <0} \gamma_\eve(\rho').  \eeqa Thus, $\gamma_\eve^{\sup}=\sup_{0 \leq \gamma_\eve <\gamma_0 } \gamma_\eve = g_\eve(\breve \rho')$, where optimal $\breve \rho'$ is given by $\breve \rho'= \arg\max_{-1 < \rho' <0} \gamma_\eve(\rho')$. Finally, we consider the existence of $\breve \rho'$ for open interval $(-1,0)$. It is straightforward to show
\beqa g_\eve (\rho') & < & \gamma_0  \\ \lim_{\rho'\rightarrow -1}g_\eve(\rho') & = & 0 \label{eq:g_2_property_1} \\ \lim_{\rho'\rightarrow 0-}g_\eve(\rho') & = & -1 \label{eq:g_2_property_2}
\\ \lim_{\rho' \rightarrow -1} \frac{\partial}{\partial \rho'}g_\eve(\rho') &=& (1+2 \ln u_\eve) u_\eve^{-1} e^R -1  \label{eq:g_2_property_3}
\eeqa where $u_\eve =\left(1-{\cal P}_\err^{\eve,\Th}\right)^{-\frac{1}{n}}  > 1$. From (\ref{eq:g_2_property_1}) and (\ref{eq:g_2_property_2}), $g_\eve(\rho')$ cannot be maximized by $\rho' \rightarrow 0-$.
If $\lim_{\rho' \rightarrow -1} \frac{\partial}{\partial \rho'}g_\eve(\rho')>0$, from (\ref{eq:g_2_property_1}) and (\ref{eq:g_2_property_2}), it is clear that $g_\eve(\rho')$ cannot be maximized by $\rho' \rightarrow -1$ and a solution must exist in $(-1,0)$. Furthermore, in this case, the maximum value $g_\eve(\breve \rho')$ must be positive due to (\ref{eq:g_2_property_1}). On the other hand, if $\lim_{\rho' \rightarrow -1} \frac{\partial}{\partial \rho'}g_\eve(\rho) =  \left((1+2 \ln u_\eve) u_\eve^{-1} e^R -1 \right) \leq 0$, we have $g_\eve(\rho') = (1+\rho')\left(u_\eve^{\frac{1}{\rho'}}e^R-1 \right)< (1+\rho')\left(u_\eve^{-1} e^R-1 \right)< (1+\rho')\left((1+2 \ln u_\eve) u_\eve^{-1} e^R-1 \right) \leq 0$ for all $-1< \rho' < 0$.
Thus, it follows from (\ref{eq:g_2_property_1}) that a solution does not exist for the open interval of $(-1,0)$.


\section*{Appendix D\\Proof of Theorem 3}
\renewcommand\theequation{D.\arabic{equation}}
\setcounter{equation}{0}
The optimization problem is
\begin{subequations} \beqa && \min_{p(\Gamma_\bob,\Gamma_\eve) } \Pr \left(\min_{0\leq \rho  \leq 1}  P_\err^{\bob,\up}(R,\rho,q_{\cal CN}(x),\Gamma_\bob, p(\Gamma_\bob,\Gamma_\eve)) >   {\cal P}_\err^{\bob,\Th}  \right) \\ && {\rm subject ~ to} ~  p(\Gamma_\bob,\Gamma_\eve) \geq 0 \\ && \hspace{1.9cm} \mathbb{E}[p(\Gamma_\bob,\Gamma_\eve)] \leq p_\av  \\ && \hspace{1.9cm} \max_{-1< \rho' \leq 0}   P_\err^{\eve,\low}(R,\rho',q_{\cal CN}(x),\Gamma_\eve, p(\Gamma_\bob,\Gamma_\eve)) \geq {\cal P}_\err^{\eve,\Th}  \label{eq:instantaneous_sec_cond} \eeqa \label{eq:power_optimization_appendix_C} \end{subequations} where $P_\err^{\bob,\up}(R,\rho,q_{\cal CN}(x),\Gamma_\bob, p(\Gamma_\bob,\Gamma_\eve))=2 P_\ber^{\bob,\up}(R,\rho,q_{\cal CN}(x),\Gamma_\bob, p(\Gamma_\bob,\Gamma_\eve))$ and  $P_\err^{\eve,\low}(R,\rho', \break q_{\cal CN}(x),\Gamma_\eve, p(\Gamma_\bob,\Gamma_\eve))= P_\ber^{\eve,\low}(R,\rho',q_{\cal CN}(x),\Gamma_\eve, p(\Gamma_\bob,\Gamma_\eve))/{\cal P}_\ber^{\rm SPN, \low}(r,K)$.
As shown in Appendix B, the interval for optimizing $\rho$ can be restricted to $0< \rho \leq 0$ and the interval for optimizing $\rho'$ can be restricted to $-1< \rho' <0$. In (\ref{eq:power_optimization_appendix_C}), $\rho$ is optimized to minimize $P_\err^{\bob, \up}(R,\rho, q_{\cal CN}(x), \Gamma_\bob, p)$. But, this is equivalent to optimizing $\rho$ to minimize the outage probability. Also, in (\ref{eq:power_optimization_appendix_C}), $\rho'$ is optimized to maximize $P_\err^{\eve, \low}(R,\rho', q_{\cal CN}(x), \Gamma_\eve, p)$, which maximizes the probability that the instantaneous security condition (\ref{eq:instantaneous_sec_cond}) is satisfied. But, this is equivalent to optimizing $\rho'$ to minimize the outage probability, because an outage is declared whenever the condition is not satisfied. Therefore, the problem of (\ref{eq:power_optimization_appendix_C}) is equivalent to the following:\begin{subequations} \beqa && \min_{0<\rho  \leq 1,-1< \rho' <0} \min_{p(\Gamma_\bob,\Gamma_\eve,\rho ,\rho') } \Pr \left( P_\err^{\bob,\up}(R,\rho,q_{\cal CN}(x),\Gamma_\bob, p(\Gamma_\bob,\Gamma_\eve,\rho ,\rho')) >   {\cal P}_\err^{\bob,\Th}  \right) \\ && {\rm subject ~ to} ~  p(\Gamma_\bob,\Gamma_\eve,\rho ,\rho') \geq 0 \\ && \hspace{1.9cm} \mathbb{E}[p(\Gamma_\bob,\Gamma_\eve,\rho ,\rho')] \leq p_\av  \\ && \hspace{1.9cm}   P_\err^{\eve,\low}(R,\rho',q_{\cal CN}(x),\Gamma_\eve, p(\Gamma_\bob,\Gamma_\eve,\rho ,\rho')) \geq {\cal P}_\err^{\eve,\Th}   \eeqa \label{eq:equivalent_optimization_1} \end{subequations} where power $p(\Gamma_\bob,\Gamma_\eve, \rho ,\rho')$ is denoted as an explicit function of $\rho$ and $\rho'$.

First, we focus on the inner optimization over $p(\Gamma_\bob,\Gamma_\eve,\rho ,\rho')$, and then we later solve the outer optimization over $\rho$ and $\rho'$.
In order to solve the inner optimization problem, following the approach of \cite{caire}, we consider the problem of minimizing power to avoid any reliability outage along with the original constraints except the total average power constraint as follows: \begin{subequations} \beqa &&  \min p(\Gamma_\bob,\Gamma_\eve,\rho,\rho') \\ && {\rm subject ~ to} ~  p(\Gamma_\bob,\Gamma_\eve,\rho ,\rho') \geq 0 \\ && \hspace{1.9cm} P_\err^{\bob,\up}(R,\rho,q_{\cal CN}(x), \Gamma_\bob, p(\Gamma_\bob,\Gamma_\eve,\rho ,\rho')) \leq   {\cal P}_\err^{\bob,\Th} \label{eq:reliability_condition_app_A} \\ && \hspace{1.9cm}  P_\err^{\eve,\low}(R,\rho',q_{\cal CN}(x), \Gamma_\eve, p(\Gamma_\bob,\Gamma_\eve,\rho ,\rho')) \geq {\cal P}_\err^{\eve,\Th}. \label{eq:security_condition_app_A} \eeqa  \end{subequations} It can be shown that, from the reliability condition of (\ref{eq:reliability_condition_app_A}),  the solution to this optimization problem must be given in the form of $p_{\min} (\Gamma_\bob,\rho) = \frac{1}{ \Gamma_\bob}g_\bob (\rho)$. Furthermore, from the security condition of  (\ref{eq:security_condition_app_A}), $p_{\min} (\Gamma_\bob,\rho)$ can be a valid solution only when the following inequality is satisfied: \beq p_{\min} (\Gamma_\bob,\rho) \leq p_{\max} (\Gamma_\bob,\rho')\eeq where $p_{\max}(\Gamma_\eve, \rho')=  \frac{1}{ \Gamma_\eve}g_\eve(\rho')$.
Then, following \cite[Proposition 4]{caire}, the optimal solution to the inner optimization problem of (\ref{eq:equivalent_optimization_1}) is given by \beq p_{\opt} (\Gamma_\bob,\Gamma_\eve, \rho, \rho') = \left\{ \begin{array}{ll} p_{\min}(\Gamma_\bob, \rho), & {\rm if ~} p_{\min}(\Gamma_\bob, \rho) \leq p_{\max}(\Gamma_\eve, \rho') {~ \rm and ~}  p_{\min}(\Gamma_\bob, \rho) \leq z_\opt
\\ 0, & {\rm if ~} p_{\min}(\Gamma_\bob, \rho) \leq p_{\max}(\Gamma_\eve, \rho')   {~ \rm and ~} p_{\min}(\Gamma_\bob, \rho) > z_\opt
\\ 0, & {\rm if ~} p_{\min}(\Gamma_\bob, \rho) > p_{\max}(\Gamma_\eve, \rho')   \end{array} \right.  \eeq
where $z_\opt$ is determined such that the average power constraint is satisfied: \beq z_\opt = \max \{z: z \geq 0, \mathbb{E}[p_{\opt}(\Gamma_\bob,\Gamma_\eve,  \rho, \rho')] \leq p_\av \}. \eeq

We now solve the outer optimization of (\ref{eq:equivalent_optimization_1}), i.e., optimizing $\rho$ and $\rho'$ to minimize the reliability outage probability:
\beqa (\breve \rho, \breve \rho') &=&  \min_{0<\rho \leq 1,-1<\rho'<0 }
\left. \Pr \left( P_\err^{\bob,\up}(R,\rho,q_{\cal CN}(x),\Gamma_\bob, p) >  \calP_\err^{\bob,\Th}  \right) \right|_{p=p_{\opt}(\Gamma_\bob,\Gamma_\eve, \rho, \rho')}.  \eeqa
Because the reliability outage occurs if and only if $p_{\opt}(\Gamma_\bob,\Gamma_\eve, \rho, \rho')=0$, we have \beqa (\breve \rho,\breve \rho') &=& \min_{0<\rho \leq 1,-1<\rho'<0 } \Pr \left( p_{\min}(\Gamma_\bob, \rho) > p_{\max}(\Gamma_\eve, \rho')  {\rm ~~or ~~} p_{\min}(\Gamma_\bob, \rho) > z_\opt \right). \eeqa
This joint optimization for $\rho$ and $\rho'$ is equivalent to two independent optimizations: $\min_{0< \rho\leq 1} p_{\min}(\Gamma_\bob, \rho)$ and $\max_{-1< \rho' < 0} p_{\max}(\Gamma_\eve, \rho')$, which are equivalent to (\ref{eq:optimization_rho_1}) and (\ref{eq:optimization_rho_prime_1}), respectively. Note that these optimizations are independent of $\Gamma_\bob$ and $\Gamma_\eve$; that is, $\breve \rho$ and $\breve \rho'$ are independent of the instantaneous channels.





\rm
\bibliographystyle{IEEE}

\end{document}